\documentclass[12pt,twoside]{article}
\usepackage{a4,amsmath,latexsym,exscale,amssymb,epsf}
\newcommand {\al}   {\alpha}       \newcommand {\bt}  {\beta}
\newcommand {\g }   {\gamma}       \newcommand {\G }  {\Gamma}
\newcommand {\dl}   {\delta}       
\newcommand {\z }   {\zeta}        
\newcommand {\s }  {\sigma}
         
\newcommand {\f }   {\varphi}      
         
\newcommand {\Lm}   {\Lambda}      \newcommand {\Om}  {\Omega}
       
\newcommand {\pl}   {\partial}     \newcommand {\nb}  {\nabla}
\newcommand {\const}  {{\sf const}} 
\newcommand {\diag}  {{\sf diag}}  
   \newcommand {\MH}  {{\mathbb H}}
\newcommand {\ML }  {{\mathbb L}}   \newcommand {\MM}  {{\mathbb M}}
\newcommand {\MR }  {{\mathbb R}}   \newcommand {\MS}  {{\mathbb S}}
\newcommand {\MT }  {{\mathbb T}}
\newcommand {\MU }  {{\mathbb U}}   \newcommand {\MV}  {{\mathbb V}}

\newcommand {\FigA} {1}
\newcommand {\FigB} {2}
\newcommand {\FigBb} {3}
\newcommand {\FigC} {4}
\newcommand {\FigD} {5}
\newcommand {\biA} {\textit{\textbf A}\/}
\newcommand {\biB} {\textit{\textbf B}\/}
\newcommand {\biC} {\textit{\textbf C}\/}

\begin{document}
\begin{titlepage}
  \renewcommand{\thefootnote}{\fnsymbol{footnote}}
\begin{center}

  \hspace*{\fill} TUW-98-09

  \vspace*{\fill}

  \textbf{\Large Global properties of warped solutions\\
                   in General Relativity}

  \vspace{7ex}

  M.~O.~Katanaev\footnotemark[1],
  T.~Kl\"osch\footnotemark[2],
  W.~Kummer\footnotemark[3]

  \vspace{7ex}

  {\footnotemark[1]\footnotesize Steklov Mathematical Institute \\
    Gubkin St. 8, Moscow 117966, Russia}

  \vspace{2ex}

  {\footnotemark[2]\footnotemark[3]\footnotesize Institut f\"ur
    Theoretische Physik \\ Technische Universit\"at Wien \\ Wiedner
    Hauptstr.  8--10, A-1040 Wien, Austria}

  \footnotetext[1]{E-mail: \texttt{katanaev@mi.ras.ru}}
  \footnotetext[2]{E-mail: \texttt{kloesch@tph.tuwien.ac.at}}
  \footnotetext[3]{E-mail: \texttt{wkummer@tph.tuwien.ac.at}}

\end{center}

\vspace{7ex}

\begin{abstract}
  Assuming the four-dimensional space-time to be a general warped product
  of two surfaces we reduce the four-dimensional Einstein equations to
  a two-dimensional problem which can be solved. All global vacuum solutions
  are explicitly constructed and analysed. The classification of the
  solutions includes the Schwarzschild, the (anti-)de Sitter, and other
  well-known solutions but also many exact ones whose detailed
  global properties to our knowledge have not been discussed before.
  They have a natural physical interpretation describing
  single or several wormholes, domain walls of curvature singularities,
  cosmic strings, cosmic strings surrounded by domain walls, solutions
  with closed timelike curves, etc.
\end{abstract}

\vspace*{\fill}
\hspace*{\fill} November 11, 1998\\
\vspace*{\fill}

\renewcommand{\thefootnote}{\arabic{footnote}}
\setcounter{footnote}{0}
\end{titlepage}

\section{Introduction}                              \label{sec:intro}
Since Einstein's discovery of general relativity the search for
and the study of exact solutions has been a field of continuous
activity. As a consequence the literature on the subject is huge.
Several examples of systematic approaches are described in
\cite{KrStMaHe80}. It is a common feature of all exact solutions that
special symmetries are assumed which lead to Killing fields on
the corresponding manifolds. The venerable Schwarzschild solution
with a spherical symmetry is the simplest example, corresponding
to an effective $2d$ theory. It may be derived by spherically
symmetric reduction from the Einstein-Hilbert action of general
relativity. A $1+1$ metric and a dilaton field are its
effective variables \cite{Thomi84}.

The increasing interest in string theories also stimulated
studies of $2d$ covariant theories. Generalizing the (trivial) $2d$
Einstein gravity from string theory, e.g., led to gravity
theory in $1+1$ dimensions with nonvanishing torsion \cite{KV86}. String
theory also inspired the dilaton black hole \cite{DBH}, whose
Lagrangian closely resembles the one of spherically reduced general
relativity. Although its global
solution is described by the same Carter-Penrose diagram as the
one for the Schwarzschild black hole, the singularity is
different, admitting complete null geodesics \cite{KAT97}. Within the
past decade also generalized dilaton theories have been studied
widely \cite{Dilrev} and, e.g., their usefulness for modelling (even
globally) the Schwarzschild black hole was shown \cite{Kat96}. Important
progress has been made in the nonperturbative quantum treatment
of such theories where the use of a ``light-cone'' gauge for the
Cartan variables leads to an Eddington-Finkelstein gauge for the
metric. This allows for an exact path integral in the geometric
variables \cite{WK92}.

Most of the considered $2d$ gravity models are simple enough to be
integrable. This raises the fundamental question of how to construct the
corresponding maximally extended surfaces or global solutions. By
this we mean that any extremal (or geodesic) can be continued to
an infinite value of its canonical parameter in both directions unless
it runs into a curvature singularity at some finite distance.
This is an important issue because almost all coordinate representations
of metrics are not maximally extended, indicating that they describe only
a small part of a larger solution. A well-known example is the
Schwarzschild solution, the maximal extension of which can be covered by
Kruskal-Szekeres coordinates \cite{Krus60}.

The Carter--Penrose diagrams for the most important Schwarzschild,
Reissner--Nordstrom, and Kerr solutions were constructed by Carter
and discuseed in his seminal review \cite{Carter73}. His method is
based on the explicit construction of overlapping coordinate systems.
It is applicable only to metrics of Lorentz signature and thus the
universality of the method is unclear. A systematic and equivalent
method for the construction of maximally extended solutions
was proposed in \cite{Katana93A} for $2d$ gravity with
torsion in the conformal gauge. It is based on the analysis of
extremals and does not use overlapping coordinate systems. The
disadvantage of the approach is that the continuity of the metric
was proved only up to second order on the horizons. A more
general approach based on a Poisson-sigma-model formulation of
generalized dilaton-type models and $2d$ gravity with
torsion has been developed in \cite{KS}. It uses overlapping coordinate
systems and proves that the maximally extended solutions are smooth.
These rules may be used for a large class of
$2d$ metrics without explicit construction of
Kruskal-Szekeres-like coordinate systems and without concern for
smoothness of the resulting maximally extended solution;
hence it has been proved in a general case. A systematic construction
of the maximally extended solutions for the Euclidean signature metrics
was proposed in \cite{Katana93A}.

Starting from Einstein's gravity in four dimensions, we
make a special ansatz for a metric reducing the problem
to a two-dimensional one, and then construct all corresponding
maximally extended solutions. We assume that the
four-dimensional space-time manifold $\MM$ with local
coordinates $x^i$ ($i,j,\dotsc =0,1,2,3$) is a warped product of two
surfaces $\MM=\MU\times \MV$. This means that in a suitable coordinate
system $x^i=\{x^\al,y^\mu\}$, ($\al,\bt,\dotsc =0,1$),
($\mu,\nu,\dotsc=2,3$), where $x^\al$ and $y^\mu$ are coordinates
on the surfaces $\MU$ and $\MV$, respectively, the four-dimensional
metric has a block diagonal form
\begin{equation}                                       \label{emetbl}
  \widehat g_{ij}=\left(
  \begin{array}{cccc}
  k(y)g_{\al\bt}(x) & 0 \\
  0 & m(x)h_{\mu\nu}(y)
  \end{array}\right) \,.
\end{equation}
Here and in what follows we use the hat symbol to denote geometrical
quantities which refer to the whole spacetime $\MM$, whereas symbols
without hats are used for $2d$ surfaces. Accordingly,
$g_{\al\bt}$ and $h_{\mu\nu}$ are metric tensors and $m(x)$ and
$k(y)$ are scalar (dilaton) fields on the surfaces $\MU$ and $\MV$,
respectively. Greek letters from the beginning ($\al,\bt,\dots$) and
the middle ($\mu,\nu,\dots$) of the alphabet always refer to
coordinates on the first and the second surface.

For definiteness we specify $\MU$ to be a pseudo-Riemannian
manifold with a Lorentz signature metric and $\MV$ a Riemannian
manifold with positive or negative definite metric. Then up to a
permutation of the first two coordinates the signature of the metric
on $\MM$ will be either $(+---)$ or $(-+++)$ depending on the sign
of $m$. These solutions are related to each other by an inversion
of the metric $g_{ij}\rightarrow-g_{ij}$ which is a symmetry
transformation of the Einstein equations in the absence of matter
fields. Both surfaces are assumed to be orientable.
Note that we do not assume initially any symmetry of the metric
(\ref{emetbl}). Only as the consequence of Einstein's equations
and the requirement of completeness we find that any solution
of the form (\ref{emetbl}) must be highly symmetric

Many well-known metrics of general relativity have this form. For
example, for the Schwarzschild solution the surface $\MV$ is a
$2d$ sphere $\MS^2$ and $\MU$ is represented by the
Carter-Penrose diagram for a black hole \cite{Carter73}.
For some recent work on warped metrics cf.\ \cite{warp}.

As far as their local properties are concerned all solutions discussed
below are well known and belong to type $D$ according to Petrov's
classification \cite{Petrov69}. Also the global structure of spherically
symmetric solutions has been discussed already in \cite{Carter73}.
The global structure of the planar and Lobachevsky plane solutions
were described quite recently \cite{HuaLia95}. The global structure
of the hyperbolic and Minkowski plane solutions constructed in the
present paper is new. The physical interpretation relies on the global
structure. We find that the solutions describe domain walls of curvature
singularities, cosmic strings surrounded by domain walls, and others.
In this construction we use the teqnique of \cite{Katana97}.

The metric (\ref{emetbl}) allows us
to solve explicitly the four-dimensional vacuum Einstein equations
with a cosmological constant $\Lm$
\begin{equation}                                       \label{einseq}
  \widehat R_{ij}=\Lm\widehat g_{ij},
\end{equation}
and to construct the possible (maximally extended) global solutions.
In Section 2 we perform the reduction of Einstein's equations for the
metric (\ref{emetbl}). The Einstein equations severely restrict the dilaton
fields; in fact, at least one of them must be constant. Hence solutions
are grouped into three main classes, corresponding to the dilaton
fields being both constant (case {\it A}), only $k = \const$ (case {\it B}),
or only $m=\const$ (case {\it C\/}).
As shown in Section~3, case {\it A\/} leads to solutions of a rather simple
type (product of two constant curvature surfaces).
Spatially symmetric solutions (case {\it B\/}) are the subject of Section~4.
Here the well-known spherically and planar solutions are
rederived. However, also a class of solutions which
are  warped products of some pseudo-Riemanian surface with a
Lobachevsky plane or with a higher genus Riemanian surface
having O(1,2) as symmetry group appear. They are physically interpreted
as $n$ wormhole solutions, where $n$ is the number of handles on
the corresponding Riemann surface. It should be noted that in this
connection that planar and Lobachevsky plane solutions of black hole
type recently attracted growing interest \cite{HuaLia95,CaiZha96}.
In Section 5 for the case {\it C\/} ($m = \const$) we have constructed new
vacuum solutions to the Einstein equations which seem to be unknown
at the global level. Again their global properties are discussed in detail.
The physical interpretation of these solutions is quite interesting.
They describe cosmic strings and domain walls of curvature
singularities. In Section 6 we summarize our results.
\section{Two-dimensional reduction}                    \label{stwodr}
The inverse metric to (\ref{emetbl}) is
\begin{equation}                                       \label{einvme}
  \widehat g^{ij}=\left(
  \begin{array}{cc}
  {\displaystyle\frac1k} g^{\al\bt} & 0 \\
  0 & {\displaystyle\frac1m} h^{\mu\nu}
  \end{array}\right),
\end{equation}
where $g^{\al\bt}$ and $h^{\mu\nu}$ are inverse matrices to $g_{\al\bt}$
and $h_{\mu\nu}$. The components of the Christoffel symbols become
\begin{eqnarray}                                       \label{echris1}
  \widehat\G_{\al\bt}{}^\g&=&\G_{\al\bt}{}^\g,
\\                                                     \label{echris2}
  \widehat\G_{\al\bt}{}^\mu&=&-\frac12g_{\al\bt}
  \frac{h^{\mu\nu}\pl_\nu k}m,
\\                                                     \label{echris3}
  \widehat\G_{\al\mu}{}^\g&=&\widehat\G_{\mu\al}{}^\g
  =\frac12\dl_\al^\g\frac{\pl_\mu k}k,
\\                                                     \label{echris4}
  \widehat\G_{\al\mu}{}^\nu&=&\widehat\G_{\mu\al}{}^\nu
  =\frac12\dl_\mu^\nu\frac{\pl_\al m}m,
\\                                                     \label{echris5}
  \widehat\G_{\mu\nu}{}^\al&=&-\frac12h_{\mu\nu}
  \frac{g^{\al\bt}\pl_\bt m}k,
\\                                                     \label{echris6}
  \widehat\G_{\mu\nu}{}^\rho&=&\G_{\mu\nu}{}^\rho.
\end{eqnarray}

With the definition of the curvature tensor
\begin{equation}                                       \label{ecurte}
  \widehat R_{ijk}{}^l=\pl_i\widehat\G_{jk}{}^l
  -\widehat\G_{ik}{}^m\widehat\G_{jm}{}^l-(i\leftrightarrow j),
\end{equation}
the components of the Ricci tensor,
$\widehat R_{ij}=\widehat R_{ikj}{}^k$, are
\begin{eqnarray}                                       \label{eritab}
  \widehat R_{\al\bt}&=&R_{\al\bt}
  +\frac{\nabla_\al\nabla_\bt m}m
  -\frac{\nabla_\al m\nabla_\bt m}{2m^2}
  +\frac{g_{\al\bt}\nabla^2 k}{2m}
\\                                                     \label{eritam}
  \widehat R_{\al\mu}&=&\widehat R_{\mu\al}
  =-\frac{\nabla_\al m\nabla_\mu k}{2mk}
\\                                                     \label{eritmn}
 \widehat R_{\mu\nu}&=&R_{\mu\nu}
  +\frac{\nabla_\mu\nabla_\nu k}k
  -\frac{\nabla_\mu k\nabla_\nu k}{2k^2}
  +\frac{h_{\mu\nu}\nabla^2 m}{2k},
\end{eqnarray}
where
\begin{equation}                                       \label{edalan}
  \nabla^2 m=g^{\al\bt}\nabla_\al\nabla_\bt m,~~~~~~
  \nabla^2 k=h^{\mu\nu}\nabla_\mu\nabla_\nu k.
\end{equation}
The covariant derivative acting on the scalar fields $m,k$ coincides with
the partial derivative while it acts on vectors on $\MU$ as
\begin{equation}                                       \label{ecodev}
  \nabla_\al V^\bt=\pl_\al V^\bt+\G_{\al\g}{}^\bt V^\g,
\end{equation}
and analogously in $\MV$.
Now the scalar curvature, $\widehat R=\widehat R_i{}^i$, becomes
\begin{equation}                                        \label{escacu}
  \widehat R=\frac1kR^g+2\frac{\nabla^2 m}{km}
  -\frac{(\nabla m)^2}{2km^2}+\frac1mR^h+2\frac{\nabla^2 k}{km}
  -\frac{(\nabla k)^2}{2k^2m},
\end{equation}
with the obvious shorthand notations
\begin{equation}                                       \label{egrsqn}
  (\nabla m)^2=g^{\al\bt}\pl_\al m\pl_\bt m,~~~~~~
  (\nabla k)^2=h^{\mu\nu}\pl_\mu k\pl_\nu k.
\end{equation}
$R^g$ and $R^h$ are scalar curvatures of the surfaces $\MU$ and $\MV$,
respectively.

Thus the Einstein equations (\ref{einseq}) for the metric (\ref{emetbl})
reduce to
\begin{eqnarray}                                       \label{eineaa}
  R_{\al\bt}+\frac{\nabla_\al\nabla_\bt m}m
  -\frac{\nabla_\al m\nabla_\bt m}{2m^2}+\frac12g_{\al\bt}
  \left(\frac{\nabla^2 k}m-2k\Lm\right)&=&0,
\\                                                     \label{einemm}
  R_{\mu\nu}+\frac{\nabla_\mu\nabla_\nu k}k
  -\frac{\nabla_\mu k\nabla_\nu k}{2k^2}+\frac12h_{\mu\nu}
  \left(\frac{\nabla^2 m}m-2m\Lm\right)&=&0,
\\                                                     \label{eineam}
  \frac{\nabla_\al m\nabla_\mu k}{mk}&=&0.
\end{eqnarray}
Equations (\ref{eineaa}), (\ref{einemm}) are rewritten in a
more suitable form by extracting the traces which yield
scalar curvatures for the surfaces
\begin{eqnarray}                                       \label{escuru}
  R^g+\frac{\nabla^2 m}m-\frac{(\nabla m)^2}{2m^2}
  +\frac{\nabla^2 k}m-2k\Lm&=&0,
\\                                                     \label{escurv}
 R^h+\frac{\nabla^2 k}k-\frac{(\nabla k)^2}{2k^2}
  +\frac{\nabla^2 m}k-2m\Lm&=&0.
\end{eqnarray}
The traceless parts of eqs.\ (\ref{eineaa}), (\ref{einemm}) multiplied
by $m$ and $k$ take the simple form
\begin{eqnarray}                                       \label{einnaa}
  \nabla_\al\nabla_\bt m
  -\frac{\nabla_\al m\nabla_\bt m}{2m}-\frac12g_{\al\bt}
  \left[\nabla^2 m-\frac{(\nabla m)^2}{2m}\right]&=&0,
\\                                                     \label{einnmm}
  \nabla_\mu\nabla_\nu k
  -\frac{\nabla_\mu k\nabla_\nu k}{2k}-\frac12h_{\mu\nu}
  \left[\nabla^2 k-\frac{(\nabla k)^2}{2k}\right]&=&0.
\end{eqnarray}
They do not contain curvature terms at all because in two dimensions the
Ricci tensor is completely defined by the scalar curvature,
$$
  R_{\al\bt}=\frac12g_{\al\bt}R^g,
$$
and has no traceless part. Note that for a surface $\MU$ the absence of a
singularity in the scalar curvature implies its absence in the full
curvature tensor.

Thus the four-dimensional vacuum Einstein equations (\ref{einseq}) for
a metric of the form (\ref{emetbl}) are equivalent to a system of
equations (\ref{eineam})--(\ref{einnmm}).
Equations (\ref{einnaa}) and (\ref{einnmm}) contain functions which
depend only on $x$ and $y$, respectively, while coordinates in the other
equations (\ref{eineam}), (\ref{escuru}), and (\ref{escurv}) are mixed.
Equation (\ref{eineam}) is very restrictive. It states that either
$k$ or $m$ or both $k$ and $m$ are constant:
\begin{equation}                                       \label{ecases}
\begin{array}{lrr}
  A: ~~& k=\const\ne0,    ~~~ & m=\const\ne0, \\
  B: ~~& k=\const\ne0,    ~~~ & \nabla_\al m\ne0, \\
  C: ~~& \nabla_\mu k\ne0,~~~ & m=\const\ne0.
\end{array}
\end{equation}
\section{Solutions with constant curvature surfaces (case \biA).
\label{scocus}}
The most symmetric warped product solutions to the reduced Einstein
equations are obtained when both $k$ and $m$ are constants.
In this case equations (\ref{eineam}), (\ref{einnaa}), and
(\ref{einnmm}) are satisfied and the scalar curvatures of both surfaces
$\MU$ and $\MV$ are constant as a consequence of eqs. (\ref{escuru}),
(\ref{escurv}) which reduce to
\begin{eqnarray}                                       \label{ecscrg}
  R^g&=&2k\Lm,
\\                                                     \label{ecscrh}
  R^h&=&2m\Lm.
\end{eqnarray}
If $\Lm=0$ then both $\MU$ and $\MV$ are surfaces of zero curvature
and $\MM$ is Minkowski space or a compactified version (cylinder,
torus) with the metric
\begin{equation}                                       \label{eminme}
  \widehat g_{ij}=\diag (+---)~~~~{\text or}~~~~
  \widehat g_{ij}=\diag (-+++).
\end{equation}

For nonzero $\Lm$ both surfaces $\MU$ and $\MV$ are of constant
nonzero curvature. These surfaces are well-known so we
give here only explicit expressions for the corresponding metrics.
If $\MU$ is a complete pseudo-Riemannian manifold of nonzero constant
curvature, $R^g=-2K=\const$, then it may be represented as the one sheet
hyperboloid, $\MH^2$, imbedded in three-dimensional Minkowski space with
the induced metric or its universal covering space
(cf.\ e.g.\ \cite{Katana93A}).
Its symmetry group is the orthogonal group $O(1,2)$.
In stereographic coordinates its metric has the standard form
\begin{equation}                                       \label{ecoclm}
  d\Phi^2=g_{\al\bt}dx^\al dx^\bt=\frac{dt^2-dx^2}
  {\left[1+\frac K4(t^2-x^2)\right]^2},
\end{equation}
where we have denoted $t=x^0$ and $x=x^1$ in order to give the metric
a more familiar appearance. In contrast to the Riemannian case the surface
is now the same for positive and negative scalar curvature $K$ but the
metric (\ref{ecoclm}) changes its sign corresponding to permutation
of the coordinates $t\leftrightarrow x$. For $K=0$ the metric
(\ref{ecoclm}) is the usual $2d$ Minkowskian metric, and
the corresponding surface is a Minkowskian plane, $\MM^2$, with the
inhomogeneous group $IO(1,1)$ as symmetry group, resp.\ a cylinder
or a torus.

The positive definite metric for a $2d$ Riemannian manifold
of constant curvature, $R^h=-2K$, in stereographic coordinates has the form
\begin{equation}                                       \label{ecocrm}
  d\Om^2=h_{\mu\nu}dy^\mu dy^\nu=\frac{dy^2+dz^2}
  {\left[1+\frac K4(y^2+z^2)\right]^2},
\end{equation}
where $y=y^1$ and $z=y^2$.
This metric differs from (\ref{ecoclm}) only in the signs. For
positive $K>0$ it corresponds to a sphere $\MS^2$. With our definition
of the curvature and Ricci tensors the scalar curvature of a
sphere with positive definite metric is negative. For $K=0$ the
metric (\ref{ecocrm}) describes the Euclidean plane $\MR^2$, or a cylinder,
or a torus. For negative $K<0$ one has the Lobachevsky (or hyperbolic)
plane $\ML^2$ or (after compactification) some higher genus Riemannian
surface. The symmetry groups for a sphere, Euclidean and Lobachevsky planes
are $O(3)$, $IO(2)$, and $O(1,2)$, respectively. For positive and negative
scalar curvatures one can always rescale coordinates in such a way
that $K=\pm 1$.

If the scalar curvatures are constant as in (\ref{ecscrg}), (\ref{ecscrh})
then the most symmetric solution for nonzero $\Lm$ which is
the warped product of two surfaces has the form
\begin{equation}                                       \label{esymef}
  ds^2=k\frac{dt^2-dx^2}{\left[1-\frac{k\Lm}4(t^2-x^2)\right]^2}
  +m\frac{dy^2+dz^2}{\left[1-\frac{m\Lm}4(y^2+z^2)\right]^2}.
\end{equation}
Rescaling the coordinates one may set $k=\pm1$, $m=\pm1$. We choose
$k=1$ and $m=-1$ in order for the metric to have the signature
$(+---)$. Then there are three essentially different solutions
corresponding to positive, zero, and negative cosmological constant:
\begin{equation}                                        \label{econcs}
\begin{array}{llll}
  \Lm<0~~& R^g=-2|\Lm|~~& R^h=+2|\Lm|~~& \MM=\MH^2\times \ML^2, \\
  \Lm=0~~& R^g=0~~      & R^h=0      ~~& \MM=\MM^2\times \MR^2, \\
  \Lm>0~~& R^g=+2|\Lm|~~& R^h=-2|\Lm|~~& \MM=\MH^2\times \MS^2.
\end{array}
\end{equation}
Although the scalar curvature for all these manifolds (as of course for any
solution of Einstein's equations (\ref{einseq})) is constant,
$\widehat R=4\Lm$, they do not coincide with the corresponding de Sitter
solutions:
each of the surfaces $\MM^2$, $\MH^2$, $\ML^2$, $\MR^2$, and $\MS^2$ has
three Killing vector fields, and the whole space-time possesses six,
whereas the de Sitter solution has ten Killing vector fields.
The above solutions are known (see, for example, \cite{KrStMaHe80}) and belong
to the $D$ type according to Petrov's classification \cite{Petrov69}.
\section{Spatially symmetric solutions (case \biB).   \label{solcog}}
The case {\it B\/} with $k=1$ describes spatially symmetric solutions
with the symmetry group $O(3)$, $IO(2)$, or $O(1,2)$ for positive,
zero, or negative scalar curvature of $\MV$, respectively.
Here we rederive the well-known spherically, planar and Lobachevsky
plane solutions. The global space-time is a warped product of $\MU$ with
$\MS^2$, $\MR^2$, or $\ML^2$, where $\MU$ is represented by a
Carter-Penrose diagram. Although these solutions are well known we review
them in detail to make the article self contained, to formulate the
general rules of global construction, and to compare those solutions with
the hyperbolic ones.

For $k=1$ the whole set of Einstein equations
(\ref{eineam})--(\ref{einnmm}) reduces to
\begin{eqnarray}                                       \label{einmaa}
  \nabla_\al\nabla_\bt m
  -\frac{\nabla_\al m\nabla_\bt m}{2m}-\frac12g_{\al\bt}
  \left[\nabla^2 m-\frac{(\nabla m)^2}{2m}\right]&=&0,
\\                                                     \label{einmmk}
  R^h+\nabla^2 m -2m\Lm&=&0,
\\                                                     \label{einscf}
  R^g+\frac{\nabla^2 m}m-\frac{(\nabla
m)^2}{2m^2}-2\Lm&=&0.
\end{eqnarray}
Equation (\ref{einmmk}) implies that the sum of two functions depending
on different coordinates equals zero. Therefore each of the functions
must be a constant. Let us fix that constant as $R^h=-2K$$=\const$.
Then eq.\ (\ref{einmmk}) is replaced by
\begin{equation}                                        \label{einrhk}
  \nabla^2 m -2(m\Lm+K)=0.
\end{equation}
Excluding the case $A$ of the previous section we proceed
requiring $\nabla_\al m\ne0$. Then (\ref{einrhk}) is a first integral
to eqs.\ (\ref{einmaa}), (\ref{einscf}). To prove this
one has to differentiate equation (\ref{einrhk}), to use the identity
$$
  \left[\nb_\al,\nb_\bt\right]V_\g=-R^g_{\al\bt\g}{}^\dl V_\dl\,,
$$
to exchange the order of covariant derivatives, and to use
(\ref{einmaa}) three times to eliminate second derivatives of $m$.
After some algebra one finally obtains equation (\ref{einscf}).
Thus only equations (\ref{einmaa}) and (\ref{einrhk}) must be solved,
(\ref{einscf}) being satisfied automatically.

We now choose a conformal gauge on $\MU$,
\begin{equation}                                       \label{ecogag}
  g_{\al\bt}dx^\al dx^\bt=2gdudv=2g(d\tau^2-d\s^2),
\end{equation}
where $g(u,v)$ is a function of two light-cone coordinates on $\MU$,
\begin{equation}                                       \label{elicoc}
  u=\tau+\s\,,~~~~~~v=\tau-\s\,.
\end{equation}
Here we denote the conformal
coordinates by Greek letters because later the transformation to
Schwarzschild coordinates will require functions $r=r(\s)$ and
$t=t(\tau)$. The four-dimensional line element takes the form
\begin{equation}                                        \label{emetko}
  ds^2=2gdudv+md\Om^2.
\end{equation}
Without loss of generality we assume here that $g>0$. Otherwise one
merely has to exchange the first two coordinates.

The Christoffel symbols for the metric (\ref{ecogag}) in conformal
gauge have only two nonvanishing components
\begin{equation}                                       \label{echsyu}
  \G_{uu}{}^u=\frac{\pl_u g}g,~~~~~~\G_{vv}{}^v=\frac{\pl_v g}g,
\end{equation}
and (\ref{einmaa}), (\ref{einrhk}) take the simple form
\begin{eqnarray}                                       \label{egcoga}
  \pl_u\pl_u m -\frac{\pl_um\pl_um}{2m}-\frac{\pl_u g\pl_um}g&=&0,
\\                                                     \label{egcogb}
  \pl_v\pl_v m -\frac{\pl_vm\pl_vm}{2m}-\frac{\pl_v g\pl_vm}g&=&0,
\\                                                     \label{egcogc}
  \frac{2\pl_u\pl_v m}g-2(m\Lm+K)&=&0.
\end{eqnarray}
Thus the full set of eqs.\ (\ref{einmaa})--(\ref{einscf}) in the
conformal gauge (\ref{ecogag}) reduces to three equations for two
unknown functions $m$ and $g$. This system of equations is
overdefined and can be integrated explicitly. The first two are
ordinary differential equations and coincide with the equations of
$2d$ gravity with torsion \cite{Katana90}. Therefore we
only sketch their integration. Dividing them by $\pl_u m$ and
$\pl_v m$, respectively, they can be integrated easily. There arise
two arbitrary functions corresponding to the invariance of
(\ref{egcoga})--(\ref{egcogc}) under the conformal transformations
$u\rightarrow u'(u)$ and $v\rightarrow v'(v)$. They may be chosen
in such a way that the functions
$m(u\pm v)$ and $g(u\pm v)$ depend simultaneously on either a
time-like or a space-like coordinate
\begin{equation}                                         \label{eindva}
  \z=\frac12(u\pm v)=\tau~~\text{or}~~\s.
\end{equation}
This means that the metric has a Killing vector field as the
consequence of equations (\ref{egcoga}) and (\ref{egcogb}). We call
these solutions homogeneous and static, respectively, although this refers
only to a specified coordinate system. This is
a generalization of the Birkhoff's theorem stating that a spherically
symmetric solution of the Einstein equations must be static. The
second and the last consequence of eqs.\ (\ref{egcoga}) and
(\ref{egcogb}) is the expression for $g$ in terms of $m$,
\begin{equation}                                       \label{erelhg}
  g=\frac{|m'|}{4\sqrt{|m|}},
\end{equation}
where the prime denotes the derivative with respect to $\z$ defined in
(\ref{eindva}). Here we have used the assumption $g>0$ and for later
convenience introduced a factor $1/4$. We are free to do this
because equations (\ref{egcoga}) and (\ref{egcogb}) define $g$ only up
to an arbitrary factor. Inserting expression (\ref{erelhg}) into
the last equation (\ref{egcogc}) we arrive at the ordinary differential
equation determining $m$,
\begin{equation}                                       \label{esecom}
  \pm m''-(m\Lm+K)\frac{|m'|}{\sqrt{|m|}}=0,
\end{equation}
where the upper and lower signs, $\pm$, correspond to homogeneous and
static solutions, respectively. Its integral depends on the sign of $m$,
\begin{equation}                                       \label{efiomw}
  \left|\frac{dm}{d\z}\right|=\mp 2W(m),
\end{equation}
where
\begin{eqnarray}                                       \label{efiomp}
  W(m)&=&-\frac13\Lm m^{3/2}-Km^{1/2}-2M,~~~~~~~~~~~~~~~~m>0,
\\                                                     \label{efiomn}
  W(m)&=&-\frac13\Lm(-m)^{3/2}+K(-m)^{1/2}-2M,~~~~~~m<0.
\end{eqnarray}
Here $M$ is an arbitrary integration constant which will be seen to
coincide with the mass in the Schwarzschild solution. The equation
for positive and negative $m$ differs only in the sign of $K$.

The line element of the solution in conformal gauge is thus
\begin{equation}                                       \label{esoecg}
  ds^2=\mp\frac{W(m)}{|m|^{1/2}}(d\tau^2-d\s^2)-md\Om^2,
\end{equation}
where $m$ depends either on $\tau$ or $\s$ through equation (\ref{efiomw}).
The sign is obtained from equation (\ref{efiomw}): The left-hand side
of the equation is always positive and the same must be true for the
right-hand side. At the same time when $m$ varies from $-\infty$ to
$+\infty$ the function $W$ may change the sign. In our case $W$ is
cubic in the square root of $m$ and may have at most three zeros which divide
$\mathbb R$ into intervals where $W$ is either positive or negative.
If $W>0$ or $W<0$ then we must choose the $+$ or $-$ sign in equations
(\ref{eindva}), (\ref{esecom}), (\ref{efiomw}), and (\ref{esoecg}). This
defines the type of the solution (static or homogeneous) on each interval.
The corresponding solution yields the conformal building block for the
construction of the Carter-Penrose diagram for a maximally extended solution.
The modulus sign in the left-hand side of equation (\ref{efiomw}) is
crucial. It means that for a given range of $m$ there are two solutions:
$m(\z)$ and $m(-\z)$. That is, all solutions in the conformal gauge
are encountered in pairs related to each other by a reflection
$\tau\rightarrow-\tau$ or $\s\rightarrow-\s$. The zeros of $W$ define
horizons
separating the blocks. The gluing procedure is unique, and for given
constants in $W$ one obtains a unique universal covering space-time.
The details of this construction are exactly the same as for
$2d$ gravity with torsion \cite{Katana93A} and are
summarized in section \ref{sclako}.

An integration constant in the equation (\ref{efiomw}) corresponds to
a shift of the coordinates $\tau$ or $\s$ and will be always set to zero.
For the scalar curvature of the surface $\MU$ one finds
(using (\ref{einscf}), (\ref{esecom}) and (\ref{efiomw}))
\begin{equation}                                       \label{escsok}
  R^g=\frac23\Lm+\frac{4M}{|m|^{3/2}}.
\end{equation}
It does not depend on $K$ or on the form of $W$ in (\ref{efiomp}),
(\ref{efiomn}), and is singular at $m=0$ if $M\ne0$. The singular
part of (\ref{escsok}) gives the eigenvalue of the singular
four-dimensional Weyl tensor \cite{LanLif62}.
\begin{equation}                                        \label{ectsqu}
  \frac1{48}\widehat C_{ijkl}\widehat C^{ijkl}
  =\left(-\frac M{|m|^{3/2}}\right).
\end{equation}
\subsection{Schwarzschild coordinates                  \label{schwac}}
At this point we have reduced the whole set of the Einstein equations to
the first order ordinary differential equation (\ref{efiomw}), the
line element being given by eq.~(\ref{esoecg}). In general, equation
(\ref{efiomw}) cannot be integrated in terms of elementary functions but
its solutions can easily be analysed qualitatively. In this way we
shall classify all global solutions in the next sections. However,
local solutions of the Einstein equations may be obtained in explicit form
without solving (\ref{efiomw}). This is achieved by using
Eddington-Finkelstein or Schwarzschild coordinates.

The transformation to Schwarzschild coordinates is very simple.
There are two cases: static and homogeneous solutions. For a static
solution the function $m(\s)$ depends on a spacelike coordinate.
Restricting ourselves to positive $m$, because of the curvature
singularity at $m=0$, and
setting $m=-r^2$, $r(\s)>0$, equation (\ref{efiomw}) becomes
\begin{equation}                                       \label{efiora}
  \left|\frac{dr}{d\s}\right|=N(r),
\end{equation}
where the right-hand side
\begin{equation}                                       \label{eprfun}
  N(r)=K-\frac{2M}r-\frac{\Lm r^2}3
\end{equation}
must be positive. This requirement defines the range of $r$ for
a given $\Lm$, $K$, and $M$.
The corresponding line element reads
$$
  ds^2=\left|\frac{dr}{d\s}\right|(d\tau^2-d\s^2)-r^2d\Om^2.
$$
Substituting $(\tau,\s)$ by $(\tau,r)$ this becomes
\begin{equation}                                       \label{elisch}
  ds^2=N(r)d\tau^2-\frac{dr^2}{N(r)}-r^2d\Om^2,~~~~~~N>0,
\end{equation}
where $N$ is given by equation (\ref{eprfun}). In the case of spherical
symmetry, that is $K=1$, it was first found by Kottler \cite{Kottle18}
and is well known to be of type $D$. For $\Lm=0$ it reduces to
the usual form of the Schwarzschild solution, $M$ being interpreted
as the mass. The form of the angular part of the line element allows one to
interpret $r$ as the radius in the spherical coordinate system. This works
only for a static branch of the solution where $N>0$, i.e.\ outside of the
horizon, $r>2M$. The timelike coordinate in the solution (\ref{elisch}) is
not restricted at all, $\tau\in(-\infty,\infty)$.

Inside the horizon
the solution is homogeneous, $m=m(\tau)$ and we set $m=-t^2$,
$t(\tau)>0$. Then equation (\ref{efiomw}) reduces to
\begin{equation}                                       \label{efiort}
  \left|\frac{dt}{d\tau}\right|=-N(t),~~~~~~N<0,
\end{equation}
where
\begin{equation}                                       \label{eprfut}
  N(t)=K-\frac{2M}t-\frac{\Lm t^2}3.
\end{equation}
The metric in Schwarzschild coordinates $(t,\s)$ reads
\begin{equation}                                       \label{elisci}
  ds^2=-\frac{dt^2}{N(t)}+N(t)d\s^2-t^2d\Om^2,~~~~~~N<0\,.
\end{equation}
The range of the timelike coordinate $t$ is defined by the inequality
$N<0$ and the singularity at $t=0$, i.e.\ $0<t<2M$.

In Schwarzschild coordinates it is not necessary to solve
equation (\ref{efiora}) or (\ref{efiort}). They simply define the
transformation to conformally flat coordinates on $\MU$.
But also in conformal coordinates we do not need the explicit solution
to equation (\ref{efiora}). To analyse the behaviour of extremals and
to construct global solutions only the behaviour of $W(m)$
near its zeros and its asymptotic behaviour as $m\rightarrow\infty$ and
$m\rightarrow0$ are needed.

The analysis of this section may be considered as a kind of uniqueness
theorem stating that among the warped metrics the Schwarzschild solution is
the only static black hole configurations in vacuum with zero cosmological
constant \cite{Israel67} (for a review, see \cite{Heusle96}). Indeed,
spherical symmetry was not assumed for the metric (\ref{emetbl}).
As a consequence of the Einstein equations (\ref{einseq}) we have
found that there is a solution with $R^h=-2K=\const$. For $K=1$ the
surface $\MV$ must be a sphere $\MS^2$. Therefore the solution
must be spherically symmetric, and one immediately obtains the
Schwarzschild solution.
\subsection{Global structure                      \label{sclako}}
Solutions of the Einstein equations for $k=1$ are parametrized
by three parameters.  The cosmological constant $\Lm$ is arbitrary.
The scalar curvature $K$ of the surface $\MV$ essentially takes three values
$K=-1,0,1$. The third parameter $M$ is a constant of motion and
may be arbitrary. For the Schwarzschild solution it is interpreted
as the mass, and a physically acceptable range is $M>0$. Nevertheless,
at least for comparison we classify solutions for all values of $M$.
The maximally extended constant curvature surface $\MV$ is a sphere,
a plane (resp.\ a cylinder or a torus) or a Lobachevsky plane
(resp.\ a higher genus Riemannian surface) for $K=1$, $0$, $-1$,
respectively. Therefore, the
classification of the global solutions reduces to the classification
of the maximally extended surfaces $\MU$ with Lorentzian metric.
These surfaces are represented globally by Carter-Penrose diagrams.

A Carter-Penrose diagram is a diffeomorphic image of a maximally extended
surface on a plane, such that the two sets of null extremals are
represented by two sets of perpendicular lines ($\pm45^\circ$)
as for the Minkowskian plane.
It has global time and space orientation. The diagram consists
of three types of conformal blocks shown in Fig.~{\FigA}. Each block
corresponds to a conformally flat metric
\begin{eqnarray}                                        \label{emetcb}
  ds^2=N(q)(d\tau^2-d\s^2),
\end{eqnarray}
where $q$ is related to $\z=\s$ or $\tau$ through
(\ref{efiora}) resp.\ (\ref{efiort}), i.e.
\begin{equation}                                        \label{ecococ}
  \left|\frac{dq}{d\z}\right|=\pm N(q).
\end{equation}
\begin{figure}[t]
 \begin{center}
 \leavevmode
 \epsfxsize 12cm \epsfbox{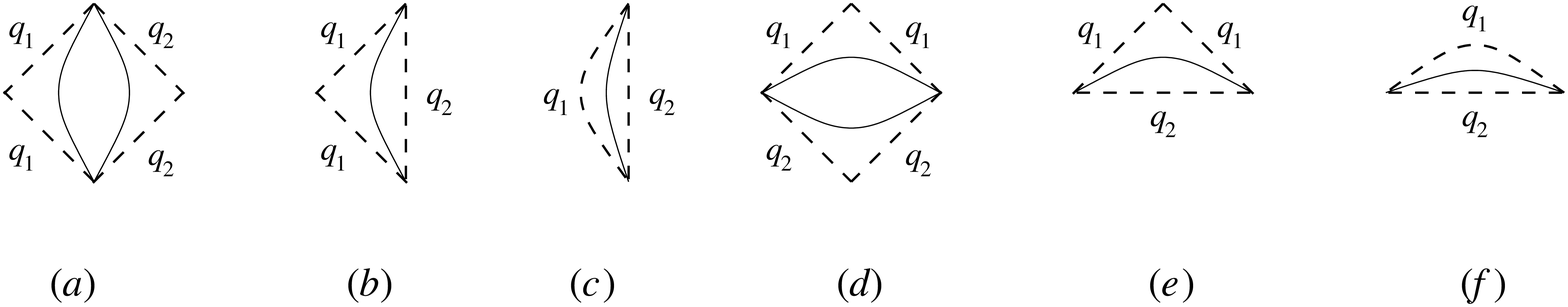}
 \end{center}
 \renewcommand{\baselinestretch}{.9}
 \small \normalsize
 \begin{quote}
 {\bf Figure \FigA:} {\small Conformal blocks for static $(a,b,c)$ and
    homogeneous $(d,e,f)$ solutions. Thin full lines indicate the direction
    of the Killing vector fields. The value of $q$
            is constant along trajectories
    and varies smoothly from the left to the right boundary for
    static solutions, being constant at the boundaries. It varies
    in a similar way but from the lower to the upper boundary for the
    homogeneous solutions. Each conformal block has its symmetric partner
    with left and right resp.\ lower and upper boundaries interchanged.}
 \end{quote}
\end{figure}
Due to (\ref{escsok}) the parameter $q$ is related to the
scalar curvature of $\MU$ by a simple algebraic relation. Thus the
metric (\ref{emetcb}) is fully defined by the scalar curvature $R^g$.
If $N>0$ or $N<0$ then the conformal block is called static or homogeneous,
respectively. We say that a static block has left and right boundary whereas
a homogeneous one has lower and upper boundary. The parameter $q$ varies
within a conformal block, being constant along the integral curves of Killing
vector fields and the boundaries. To each conformal block corresponds a
time or space reversed symmetric partner. The range of
the coordinates $\s$ and $\tau$ is determined by the zeros of the conformal
factor, $N(q_k)=0$, called horizons, and the two boundary values $q_0$ and
$q_\infty$ of the maximally extended surface
corresponding either to a singularity or to a complete boundary where
the integral (metric distance)
\begin{equation}                                         \label{eintco}
  \int ds=\int\limits^{q_\infty}\frac{dq}{\sqrt{|N(q)|}}
\end{equation}
diverges. In our case $q_\infty=\infty$, but in general the infinite boundary
with respect to the metric (\ref{emetcb}) may correspond to a finite value
of $q$. The singularity may lie either at a finite or an infinite distance
depending on the integral (\ref{eintco}). There are three types of
conformal blocks: diamond, triangle, and ``eye'' shown in Fig.~{\FigA}.
If the coordinates $\s,\tau$ cover the whole plane, then the solution
(\ref{emetcb}) is mapped onto a finite diamond, Fig.~{\FigA}~$(a)$ and $(d)$,
by the conformal transformation.
If the coordinates cover a half plane, i.e.\ the integral
\begin{equation}                                         \label{eforcb}
  \int d\z =\int\limits^{q_0,~q_k~\text{or}~q_\infty}\frac{dq}{N(q)}
\end{equation}
converges at one end (boundary value or horizon) of the respective
$q$-interval, then the solution is represented by a triangular
conformal block, Fig.{\FigA}~$(b)$ and $(e)$. If the function $N$ does not
have any zero and the integral (\ref{eintco}) converges both at $q_0$ and
$q_\infty$, then the maximally extended surface is given by an ``eye''
Carter-Penrose diagram shown in Fig.{\FigA}~$(c)$ and $(f)$. Now we give the
rules for construction of a maximally extended solution for a given
metric of the form (\ref{emetcb}) \cite{Katana93A}:
\begin{itemize}
\item Consider a range of the parameter $q$ between two boundary values
$q_0$ and $q_\infty$, where the curvature is singular or the integral
(\ref{eintco}) diverges. Call the corresponding boundaries singular and
infinite, respectively.
\item Find roots of the equation $N(q_k)=0$, $k=1,\dots,n$ between $q_0$
and $q_\infty$ and call them horizons, the degree of the zero $q_k$ being
the degree of the corresponding horizon.
\item For each of the intervals $(q_0,q_1),\dots,(q_n,q_\infty)$ draw a
pair of static or homogeneous conformal blocks for $N>0$ and $N<0$,
respectively.
\item If there are horizons, then glue conformal blocks along
horizons preserving the smoothness of $N$. That is, the boundaries
of conformal blocks must be glued together, corresponding only to the
adjacent intervals having $q_k$ as a boundary point.
\item For a given $N$ the Carter-Penrose diagram constructed by gluing
all adjacent but different
conformal blocks constitutes a fundamental
region. If $N\ge0$ or $N\le0$ everywhere between $q_0$ and $q_\infty$
(that is all possible horizons are of even degree), then there are two
disconnected fundamental regions related by space or time reflection.
\item If the boundary of the fundamental region consists of the boundaries
of the conformal blocks corresponding to $q_0$ or $q_\infty$ then the
fundamental region represents the unique smooth maximally extended surface.
\item If the boundary of the fundamental region includes horizons then
the fundamental region may be continued periodically in time or space
or both directions simultaneously, or the opposite sides may be
identified.
\end{itemize}
The final step is to determine the completeness or incompleteness of the
singular boundary by considering the integral (\ref{eintco}) at $q_0$.
There are curious situations. For example, in ordinary dilaton gravity
the singularity is complete for null extremals but incomplete
for time-like ones \cite{KAT97}. The inner points where horizons cross
may be at a finite or infinite distance depending on whether the integral
(\ref{eintco}) converges or diverges at $q_k$, i.e.\ they are at an infinite
distance iff $q_k$ is a zero of degree $\ge 2$.
Such infinite points inside the diagram do occur in $2d$ gravity
with torsion \cite{Katana93A}. The present case is simpler, and
all points inside a diagram are finite.

Near a horizon the function $N$ behaves like $(q-q_k)^a$. For $a\ge1$
the integral (\ref{eforcb}) diverges. Therefore conformal blocks between
such horizons are of diamond type.
If horizons are absent then the metric (\ref{emetcb}) covers the whole
maximally extended space-time and is represented by one conformal block.

The described procedure is quite general, unique and yields a globally
smooth maximally extended solution. The advantage of this constructive
approach is that the global structure of the space-time is defined
without explicit construction of a global coordinate system.
The starting point is the local form of the metric admitting one Killing
vector field. The global ${\cal C}^\infty$ smoothness can be proved
by explicit construction of the Eddington--Finkelstein and Kruskal--Szekeres
coordinates covering the fundamental region.
In our present paper we work with the conformal approach, the main reason
being that we have to treat Euclidean metrics as well, where the
Eddington-Finkelstein coordinates do not exist.

Thus the classification of global solutions to the Einstein equations
reduces to the analysis of the zeros of $N$ for different values of $\Lm$,
$K$, and $M$. The function $qN(q)$ is cubic in $q$ and may have up to
three zeros. One of the zeros is necessarily negative and should be
dropped. Thus, there occur at most two horizons. Below
we shall write down explicitly the local solutions of Einstein's equations
in Schwarzschild coordinates and in the conformal gauge where equation
(\ref{ecococ}) can be solved using elementary functions. For $\Lm=0$,
$K=0$, and $M=0$ we have $m=\const$, and the space-time is Minkowskian,
as found in Section \ref{scocus}. Hence we assume that at least one
parameter differs from zero. We classify the solutions according to the
scalar curvature of the $\MV$ surface.
\subsection{Spherically symmetric solutions, $K=1$     \label{sphers}}
For $K=1$ the surface $\MV=\MS^2$ is a sphere and all solutions are
spherically symmetric. For the unit sphere the
metric (\ref{ecocrm}) may be rewritten in spherical coordinates
\begin{equation}                                       \label{eliesp}
  d\Om^2=d\theta^2+\sin^2\theta d\f^2.
\end{equation}
The solutions are
parametrized by a cosmological constant $\Lm$ and a mass $M$. The
line element in Schwarzschild coordinates has the usual form
\begin{equation}                                        \label{esphsi}
  ds^2=N(q)d\z^2-\frac{dq^2}{N(q)}-q^2(d\theta^2+\sin^2\theta d\f^2),
\end{equation}
where
\begin{equation}                                        \label{edefus}
  N(q)=1-\frac{2M}q-\frac{\Lm q^2}3.
\end{equation}
Depending on the sign of $N$, the coordinates $q$ and $\z$ may be space-
or timelike
\begin{equation}                                        \label{ecoran}
\begin{array}{lll}
  N(q)>0, &~~~~q=r, &~~~~\z=\tau,
\\
  N(q)<0, &~~~~q=t, &~~~~\z=\s.
\end{array}
\end{equation}
The parameter $q$ must be positive. Despite the similar local form
of the solution for different values of $\Lm$ and $M$, the global
structure varies qualitatively. It depends  on the number and types of
the roots of the equation $N(q)=0$ or
\begin{equation}                                        \label{erootn}
  \frac\Lm3q^3-q+2M=0.
\end{equation}
For nonzero cosmological constant (\ref{erootn}) has up
to three zeros. Elementary analysis shows that one root is always
negative and we have at most two positive roots and, consequently,
two horizons. For positive cosmological constant, $\Lm>0$, we have the
following possibilities:
For $3M>\frac1{\sqrt\Lm}$ equation (\ref{erootn}) has no positive
root. If $3M=\frac1{\sqrt\Lm}$, then there is one double positive root.
In the interval $0<3M<\frac1{\sqrt\Lm}$ there are two positive roots.
For $M\le0$ we have one positive root. For negative cosmological
constant, $\Lm<0$, there is one positive root for $M>0$ and no positive
root for $M<0$.
\subsubsection{Minkowskian space-time, $\Lm=0$, $M=0$}
The simplest spherically symmetric solution is obtained for $\Lm=0$
and $M=0$ (cf.~Fig.~{\FigB}~$(a)$).
Then $N=1$, and the line element in the Schwarzschild coordinates becomes
\begin{equation}                                        \label{emilis}
  ds^2=d\tau^2-dr^2-r^2(d\theta^2+\sin^2\theta d\f^2),~~~~~~0<r<\infty \, .
\end{equation}
The point $r=0$ is a coordinate singularity.
Going to four-dimensional Cartesian coordinates and adding
the time developed origin, $r=0$, we get the four-dimensional
Minkowskian space-time. Here the spacelike coordinate $r$ is naturally
identified with the radius of the spherical coordinate system.
\subsubsection{Schwarzschild black hole solution, $\Lm=0$, $M>0$}
The Schwarzschild solution corresponds to zero cosmological
constant, $\Lm=0$, and positive mass, $M>0$. Its Carter-Penrose diagram
is well-known. Nevertheless, it represents a suitable non-trivial
illustration of our general approach. In this case the
right-hand side of the equation (\ref{efiora})
\begin{equation}                                        \label{eqschw}
  \left|\frac{dr}{d\s}\right|=1-\frac{2M}r,
\end{equation}
has one zero at $r_1=2M$, so there is one horizon. For definiteness
we assume here that $dr/d\s>0$ and then add solutions with $dr/d\s<0$
by reflection $\s\rightarrow-\s$. In the region $N>0$ i.e.\ $r>2M$ the
solution is static. For $r=r(\s)$ equation (\ref{eqschw}) yields
\begin{equation}                                        \label{eshwrx}
  r+2M\ln(r-2M)=\s,~~~~~~-\infty<\s<\infty \,,
\end{equation}
revealing $\s$ as the ``tortoise coordinate''.
The corresponding line element in the conformal gauge has the form
\begin{equation}                                       \label{eschsc}
  ds^2=\left(1-{\frac{2M}r}\right)(d\tau^2-d\s^2)-r^2d\Om^2,
\end{equation}
where $r(\s)$ is given implicitly by (\ref{eshwrx}),
or in Schwarzschild coordinates
\begin{equation}                                       \label{eschss}
  ds^2=\left(1-{\frac{2M}r}\right)d\tau^2
       -\frac{dr^2}{1-\frac{2M}r}-r^2d\Om^2,~~~~~~2M<r<\infty \,.
\end{equation}
The corresponding conformal blocks of a Carter-Penrose diagram are the
diamonds {\bf I}, {\bf III} shown in Fig.~{\FigB}~$(b_1)$.
\begin{figure}[t]
 \begin{center}
 \leavevmode
 \epsfxsize 13cm \epsfbox{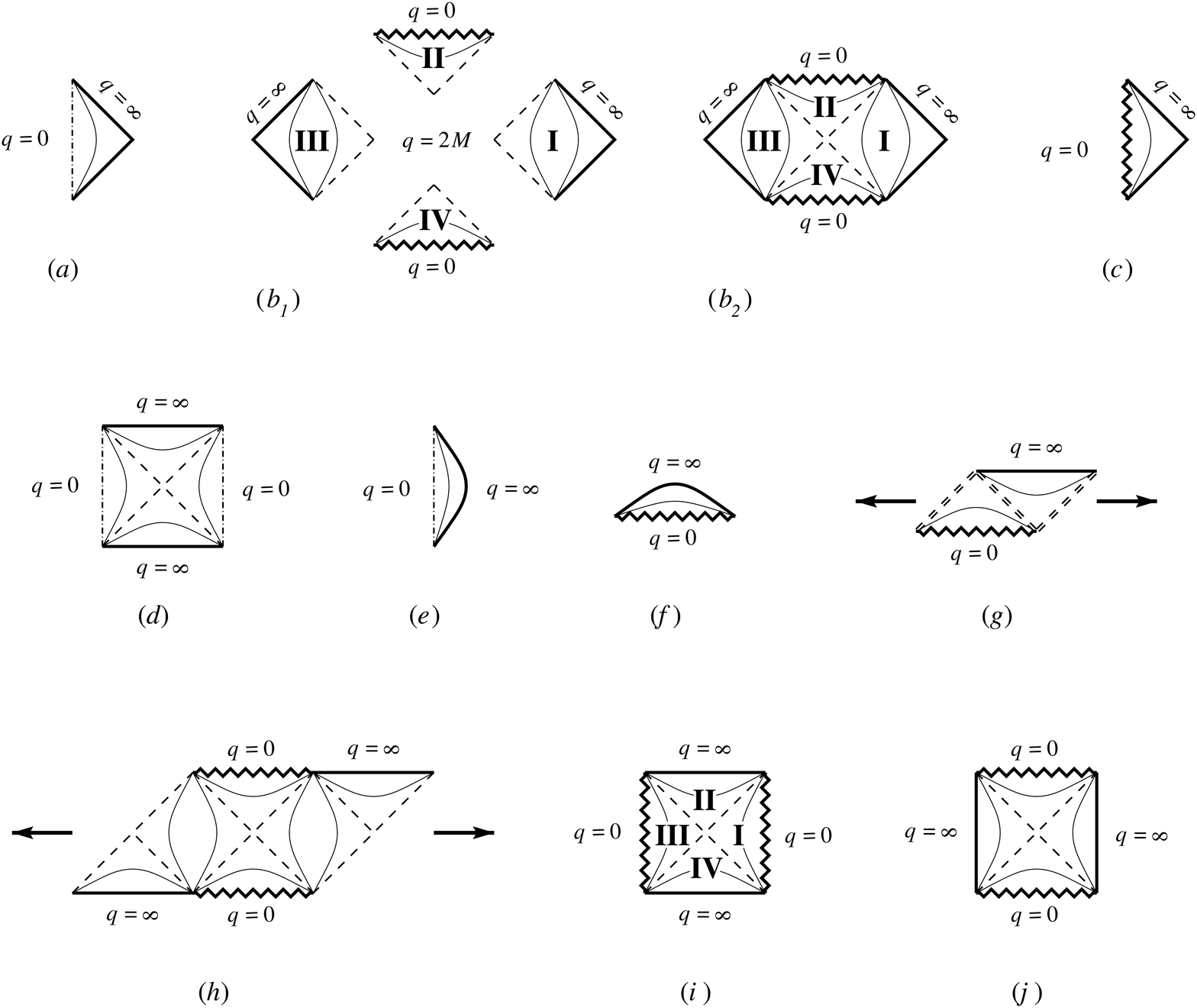}
 \end{center}
 \renewcommand{\baselinestretch}{.9}
 \small \normalsize
 \begin{quote}
 {\bf Figure \FigB:} {\small Spherically symmetric solutions, $K=1$.
       In our notations the zigzag lines denote the singular boundary
       lying at a finite distance, solid lines denote a regular
       infinite boundary, thin lines are the Killing trajectories,
       dashed lines denote horizons. Solid arrows indicate a possible
       periodic continuation of the fundamental regions.}
 \end{quote}
\end{figure}
In the region $N<0$ or $0<t<2M$ the solution depends on the time
$t=t(\tau)$, and we have to change the sign of the right-hand side of
equation (\ref{eqschw}). This implicitly defines $t(\tau)$:
\begin{equation}                                        \label{eshwrt}
  t+2M\ln(2M-t)=-\tau,~~~~~~-2M\ln(2M)<\tau<\infty \,.
\end{equation}
In conformal gauge resp.\ in Schwarzschild coordinates the line element reads
\begin{eqnarray}                                       \label{eschhc}
  ds^2&=&-\left(1-{\frac{2M}t}\right)(d\tau^2-d\s^2)-t^2d\Om^2,
\\                                                     \label{eschst}
  ds^2&=&-\frac{dt^2}{1-\frac{2M}t}
       +\left(1-{\frac{2M}t}\right)d\s^2
       -t^2d\Om^2,~~~~~~0<t<2M \,.
\end{eqnarray}
The corresponding conformal blocks are now a triangle and its reflection as
shown in Fig.~{\FigB}~$(b_1)$ {\bf II}, {\bf IV}. The corresponding
Carter-Penrose diagram for the
global solution is unique and shown in Fig.~{\FigB}~$(b_2)$.
\subsubsection{Naked singularity, $\Lm=0$, $M<0$}
For negative mass the right-hand side of equation (\ref{eqschw})
is positive for $r>0$ and the solution depends on a spacelike
coordinate $\s$. The line element has the same form (\ref{eschsc}) as for
the Schwarzschild solution. The only difference is the range of
$\s\in(2M\ln|2M|,\infty)$. Thus there is only one triangular
conformal block shown in Fig.~{\FigB}~$(c)$ and its space reflection.
Each of them represents the maximally extended surface $\MU$.
\subsubsection{De Sitter solution, $\Lm>0$, $M=0$}
The de Sitter solution corresponds to a positive cosmological constant
and zero mass. This complete constant curvature manifold may also be
represented as the ``unit'' hyperboloid embedded in 5-dimensional Minkowski
space with the induced metric. Consequently, its symmetry group equals
$O(1,4)$, and it has the maximal number of 10 Killing fields.
The function (\ref{edefus})
has one zero so there is one horizon. The static
and homogeneous solutions in Schwarzschild coordinates become
\begin{eqnarray}                                        \label{edesss}
  ds^2&=&\left(1-{\frac\Lm3}r^2\right)d\tau^2
  -\frac{dr^2}{1-\frac\Lm3r^2}-r^2d\Om^2,
\\                                                      \nonumber
  &&0<r<{\sqrt{\textstyle\frac3\Lm}}\,,~~-\infty<\tau<\infty\,,
\\                                                      \label{edeshs}
  ds^2&=&-\frac{dt^2}{1-\frac\Lm3t^2}
  +\left(1-{\frac\Lm3}t^2\right)d\s^2-t^2d\Om^2,
\\                                                      \nonumber
  &&{\sqrt{\textstyle\frac3\Lm}}<t<\infty\,,~~-\infty<\s<\infty \,.
\end{eqnarray}
Since \ref{ecococ}) can be integrated in that case, the line elements
may be written explicitly in conformal gauge, too,
\begin{eqnarray}                                        \label{edescs}
  ds^2&=&\frac1{\cosh^2\left(\sqrt{\frac\Lm3}\s\right)}(d\tau^2-d\s^2)
  -\frac3\Lm\tanh^2\left({\textstyle\sqrt{\frac\Lm3}}\s\right)d\Om^2,
\\                                                      \label{edesch}
  ds^2&=&\frac1{\sinh^2\left(\sqrt{\frac\Lm3}\tau\right)}(d\tau^2-d\s^2)
  -\frac3\Lm\coth^2\left({\textstyle\sqrt{\frac\Lm3}}\tau\right)d\Om^2,
\end{eqnarray}
for static and homogeneous solutions, respectively. The range
$r\in(0,\sqrt{3/\Lm})$ is transformed to $\s\in(0,\infty)$ and
$t\in(\sqrt{3/\Lm},\infty)$ to $\tau\in(0,\infty)$. Both line elements
(\ref{edescs}) and (\ref{edesch}) describe triangular conformal blocks,
and the Carter-Penrose diagram is shown in Fig.~{\FigB}~$(d)$.
Adding the line $r=0$ (i.e.\ the time-evolved coordinate origin)
to the warped product $\MU\times\MS^2$, the de Sitter solution is obtained.
\subsubsection{Anti de Sitter solution, $\Lm<0$, $M=0$}
Changing the sign of the cosmological constant in the Einstein
equations results in a qualitatively different solution. For zero mass
we have the anti de Sitter solution which may also be represented as
a ``unit'' hyperboloid, but this time embedded in a flat five-dimensional
space of signature $(++---)$.
The corresponding symmetry group is $O(2,3)$.
 The function (\ref{edefus}) has
no zeros and is always positive. Therefore the solution is static and
has no horizon. In Schwarzschild coordinates
the line element has the same form as for the de Sitter solution
(\ref{edesss}) but due to the minus sign of $\Lm$ the range of $r$ is
now $(0,\infty)$. In the conformal gauge we obtain
\begin{equation}                                       \label{eadscg}
  ds^2=\frac1{\cos^2\left(\sqrt{\frac{|\Lm|}3}\s\right)}(d\tau^2-d\s^2)
  -\frac3{|\Lm|}\tan^2\left({\textstyle\sqrt{\frac{|\Lm|}3}}\s\right)d\Om^2.
\end{equation}
Here $\s$ runs through the finite interval
$\s\in(0,\frac\pi2\sqrt{\frac3{|\Lm|}})$. In this case we have an ``eye''
diagram shown in Fig.~{\FigB}~$(e)$, but we could equally well draw it
as an infinite ribbon with vertical boundary lines.
This manifold is incomplete at $r=0$. To obtain the entire anti de Sitter
solution we have to add this line (time-developed point) to the manifold.
\subsubsection{Homogeneous space singularity,
               $\Lm>0$, $3M>\frac1{\protect\sqrt{\Lm}}$}
For these values of the constants the function $N$ has no zero for $q>0$
and is always negative. The solution is homogeneous without horizons.
The surface $\MU$ in that case is represented by an ``eye'' Penrose diagram
shown in Fig.~{\FigB}~$(f)$ and its time reversal. At a finite past we
observe a true singularity both in four-dimensional curvature and the
curvature of the surface $\MU$, and the manifold cannot be extended
through it.
A cosmological interpretation of this diagram implies a Universe born at
a finite past lasting forever.
The surface $\MU$ is asymptotically de Sitter as $t\rightarrow\infty$.
The same is also true for the three cases described below.
\subsubsection{One double horizon,
               $\Lm>0$, $3M=\frac1{\protect\sqrt{\Lm}}$}
In that case the function $N$ has one double zero at $t=1/\sqrt\Lm$
corresponding to a horizon, and the function $N$ is everywhere negative
for $q>0$. The Carter-Penrose diagram is the infinite ribbon shown in
Fig.~{\FigB}~$(g)$ and its time reversal.
\subsubsection{Two horizons, $\Lm>0$, $0<3M<\frac1{\protect\sqrt{\Lm}}$}
The function $N$ has two zeros (two horizons).
The Carter-Penrose diagram
consists of two homogeneous and one static conformal block. The
corresponding universal covering space is an infinite ribbon shown in
Fig.~{\FigB}~$(h)$.
\subsubsection{Two static singularities $\Lm>0$, $M<0$}
For negative mass the function $N$ has one zero. The maximally extended
solution is shown in Fig.~{\FigB}~$(i)$. We have two static singular
regions {\bf I,III} separated by horizons. From {\bf I} only the right-hand
singularity can be reached by a causal path and the same applies to
{\bf III} and the left-hand singularity.
Region {\bf IV} has access to  both (naked) singularities, whereas from
{\bf II} no singularity can be reached.
\subsubsection{Anti de Sitter black hole $\Lm<0$, $M>0$}
In that case the function $N$ has one zero, that is, one horizon.
The Carter-Penrose diagram is shown in Fig.~{\FigB}~$(j)$, and is similar
to the one for the Schwarzschild black hole except that
the complete left and right boundaries are now timelike. The singularities
at finite past and finite future are similar to those of the Schwarzschild
solution. The space-time is not asymptotically flat but becomes
anti de Sitter as $r\rightarrow\infty$.
\subsubsection{ $\Lm<0$, $M<0$}
This is a naked singularity without horizon. Its Carter-Penrose diagram
equals the one in Fig.~{\FigB}~$(f)$ but turned by $\pi/2$ as the solution
is static. The space at $r\rightarrow\infty$ is asymptotically anti de
Sitter again.
\subsection{Planar solutions, $K=0$                    \label{spherp}}
In the case $K=0$ the line element (\ref{ecocrm}) is
\begin{equation}                                        \label{evplso}
  d\Om_P^2=dy^2+dz^2
\end{equation}
and describes the Euclidean plane with the symmetry group $IO(2)$ or
its compactified versions: a cylinder or a torus. In the planar case
the line element in Schwarzschild coordinates is
\begin{equation}                                        \label{eplsoi}
  ds^2=N(q)d\z^2-\frac{dq^2}{N(q)}-q^2(dy^2+dz^2),
\end{equation}
where
\begin{equation}                                        \label{edefut}
  N(q)=-\frac{2M}q-\frac{\Lm q^2}3.
\end{equation}
Coordinates $q$ and $\z$ are the same as in equation (\ref{ecoran}).
The global solutions depend on $M$ and $\Lm$.

The physical interpretation of these solutions is quite interesting.
For instance, the whole three-dimensional space of the corresponding
solutions may be a product of a torus $\MT^2$ with a line $\MR$ (for
the choice $\MV = \MT^2$). This space contains non-contractible
loops (nontrivial fundamental group) and may be interpreted as a wormhole.
Note that the horizon in that case will be a torus too.
\subsubsection{Homogeneous and naked singularities $\Lm=0$, $M\ne0$}
%
\begin{figure}[t]
 \begin{center}
 \leavevmode
 \epsfxsize 11cm \epsfbox{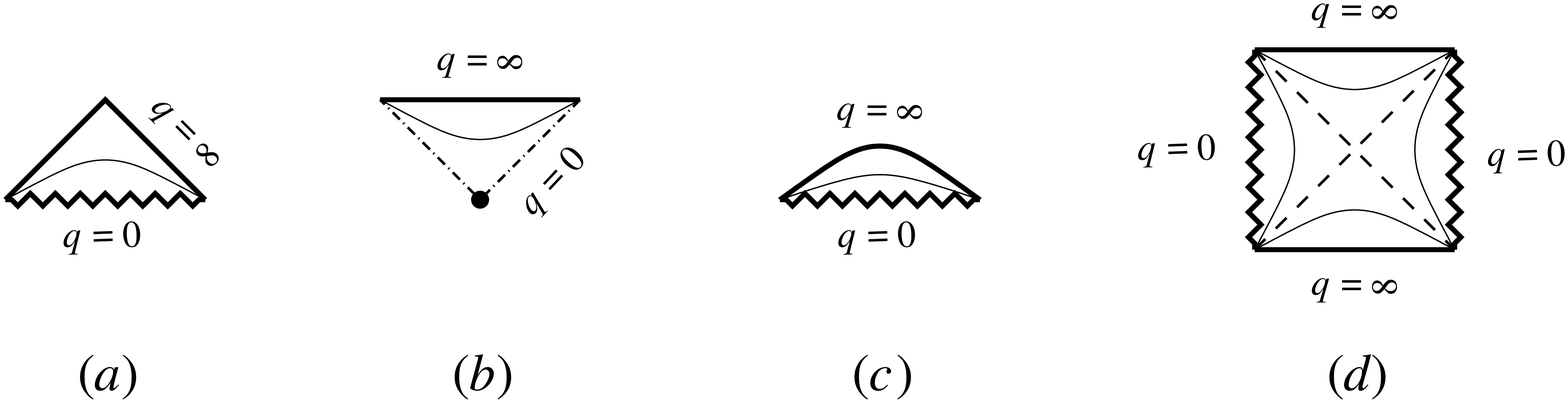}
 \end{center}
 \renewcommand{\baselinestretch}{.9}
 \small \normalsize
 \begin{quote}
 {\bf Figure \FigBb:} {\small Planar solutions, $K=0$.}
 \end{quote}
\end{figure}
For zero cosmological constant and positive mass $M>0$ we have a
homogeneous spacelike singularity. In that case $N$ is everywhere negative,
and the solution does not have a horizon. In the Schwarzschild and
conformal coordinates resp.\ the line elements are
\begin{eqnarray}                                        \label{eplmas}
  ds^2&=&\frac t{2M}dt^2-\frac{2M}td\s^2-t^2{d\Om_P}^2\,,
  ~~~~~~~~~~~~~0<t<\infty\,,
\\
  ds^2&=&\sqrt{\textstyle\frac M{|\tau|}}(d\tau^2-d\s^2)-4M|\tau|d\Om_P^2\,,
  ~~~~~~\mbox{$\tau<0$\quad or\quad $\tau>0$}\,.
\end{eqnarray}
The corresponding Carter-Penrose diagram is shown in Fig.~{\FigBb}~$(a)$.
As usual, there is also another time reversed solution.

For negative mass, $M<0$, the solution has a similar form but is stationary.
Then the Carter-Penrose diagram should be turned by $\pi/2$ and describes a
naked singularity.
\subsubsection{(Anti) de Sitter solution, $\Lm\ne0$, $M=0$}
For positive cosmological constant ($\Lm>0$) and zero mass the
function
$$
  N=-\frac\Lm3t^2
$$
is everywhere negative and has no zeros. The corresponding solution is
homogeneous without horizon. In the Schwarzschild and conformal
coordinates it reads
\begin{eqnarray}                                        \label{enlzms}
  ds^2&=&\frac3{\Lm t^2}dt^2-\frac{\Lm t^2}3d\s^2-t^2d\Om_P^2,
\\                                                      \label{enlzmc}
  ds^2&=&\frac3{\Lm\tau^2}(d\tau^2-d\s^2)-\frac9{\Lm^2\tau^2}d\Om_P^2.
\end{eqnarray}
The last equation shows that this solution is conformally flat
Along the boundary lines $q=0$ (i.e., $t=0$) of the Carter-Penrose diagram
(Fig.~{\FigBb}~$(b)$) our $4d$-metric becomes degenerate.
One can verify that metrics (\ref{enlzms}) and (\ref{enlzmc}) describe
space-time of constant curvature. This means that these metrics represent
(part of) the de Sitter solution.

The case of a negative cosmological constant, $\Lm<0$, is similar but
one has to replace the time dependence of metric components in
eqs.~(\ref{enlzms}) and (\ref{enlzmc}) by a space dependence and turn
the Carter-Penrose diagram by $\pi/2$. This is one of the forms of
anti de Sitter space-time.
\subsubsection{ $\Lm\ne0$, $M\ne0$}
If $\Lm>0$ and $M>0$ the function
\begin{equation}                                        \label{enfuzk}
  N=-\frac{\Lm t^3+6M}{3t}<0
\end{equation}
is negative without zeros. In Schwarzschild resp.\ conformal coordinates the
line elements are
\begin{eqnarray}                                        \label{epslms}
  ds^2&=&\frac{3t}{\Lm t^3+6M}dt^2-\frac{\Lm t^3+6M}{3t}-t^2d\Om_P^2\,,
  ~~~~~~t\in(0,\infty)\,.                                               \\
  ds^2&=&\left|\frac{dt}{d\tau}\right|(d\tau^2-d\s^2)-t^2d\Om_P^2\,,
  ~~~~~~ -\frac\pi{12\sqrt3M}<\tau<\frac\pi{4\sqrt3M}\,.
\end{eqnarray}
Here the function $t=t(\tau)$ is defined implicitly by the relation
\begin{equation}                                        \label{erects}
  \frac\Lm3\tau=\frac1{6a}\ln\frac{a^2-at+t^2}{(a+t)^2}
  +\frac1{a\sqrt3}\arctan\left(\frac{2t-a}{a\sqrt3}\right),
  ~~~~~~a=6M/\Lm.
\end{equation}
Correspondingly, the
inextendible solution is represented by an ``eye'' diagram as shown in
Fig.~{\FigBb}~$(c)$, and its time reversal. It describes a homogeneous space
singularity at $t=0$ and is asymptotically de Sitter as $t\rightarrow\infty$.

If $\Lm>0$, $M<0$, the function (\ref{enfuzk}) has one simple zero
corresponding to a horizon. There is one static and one homogeneous
conformal block (and their reflections) forming together the
Carter-Penrose diagram shown in Fig.~{\FigBb}~$(d)$.
It coincides with the
diagram for the spherically symmetric case $K=1$, $\Lm>0$, and $M<0$
(Fig.~{\FigB}~$(i)$).
The line element for the homogeneous block has the same form as
(\ref{epslms})but due to (\ref{erects})
the ranges of $t$ and $\tau$ differ.
For the static block the line element is given by (\ref{elisch}) with the
appropriate ranges for $r$ and $\s$ related by the same  equation as
$t$ and $\tau$.

The cases $\Lm<0$, $M<0$ and $\Lm<0$, $M>0$ are obtained from the
cases $\Lm>0$, $M>0$ and $\Lm>0$, $M<0$ by exchanging timelike and
spacelike coordinates. The last case corresponds to a torus black hole.
\subsection{Lobachevsky plane solutions, $K=-1$        \label{sootss}}
For $K=-1$ the surface $\MV$ is a Lobachevsky plane or, in a compactified
version, a higher genus Riemannian surface. Note that the Lorentz
transformation group $O(1,2)$ acts on a spacelike surface in that case.
The line element (\ref{ecocrm}) in angular coordinates is
\begin{equation}                                       \label{eilpac}
  d\Om_L^2=d\theta^2+\sinh^2\theta d\f^2.
\end{equation}
The corresponding vacuum solution to the Einstein equation has the form
\begin{equation}                                        \label{elosoi}
  ds^2=N(q)d\z^2-\frac{dq^2}{N(q)}-q^2(d\theta^2+\sinh^2\theta d\f^2),
\end{equation}
where
\begin{equation}                                        \label{edelos}
  N(q)=-1-\frac{2M}q-\frac{\Lm q^2}3.
\end{equation}
The function $W$ as given by (\ref{efiomn}) changes its overall sign
under the reflection $K\rightarrow-K$, $\Lm\rightarrow-\Lm$, and
$M\rightarrow-M$. Thus the classification of solutions for $K=-1$
is the same as for the spherically symmetric case if one interchanges
space and time coordinates on the $\MU$ surface, $\tau\leftrightarrow\s$,
and changes the sign of cosmological constant and mass. For example,
the analog of the Schwarzschild solution is obtained
for $\Lm=0$ and $M<0$.
For homogeneous and static conformal blocks the line element is
\begin{equation}                                       \label{eschsf}
  ds^2=\frac{dt^2}{1+\frac{2M}t}
       -\left(1+{\textstyle\frac{2M}t}\right)d\s^2
       -t^2d\Om^2,~~~~~~-2M<t<\infty \,,
\end{equation}
\begin{equation}                                       \label{eschse}
  ds^2=\left(1+{\textstyle\frac{2M}r}\right)d\tau^2
       -\frac{dr^2}{1+\frac{2M}r}-r^2d\Om^2,~~~~~~0<r<-2M \,.
\end{equation}
The corresponding Carter-Penrose diagram must be turned by the
angle $\pi/2$. There are left and right static singularities at $r=0$.
The properties of this global solution are similar to the properties of
two static singularities in the spherically symmetric case, but
with a different ``asymptotic'' behaviour: the space-time is now
asymptotically flat at infinite past and infinite future (both limits
correspond to $t\rightarrow\infty$).

It is not easy to imagine a space (a time slice of the four-dimensional
space-time defined by the equation $t=\const$) for $K=-1$. In an
uncompactified version it is the product of an interval (finite or infinite
depending on the values of $\Lm$ and $M$) and the Lobachevsky plane. In a
compactified version when $\MV$ is a higher genus Riemann surface
one may interpret this class of solutions as describing a set of
wormholes, the number of wormholes being exactly the number of handles
of the corresponding Riemann surface. The same topology will be inherited
by a horizon. In the same way one may analyse the other solutions for
$K=-1$, but we shall not repeat this analysis here.
\section{Hyperbolic solutions (case {\biC}).          \label{solcos}}
Whereas solutions of the last section are known, the global properties
of the solutions to be discussed now seem to be new. Their topological
structure is $\MM=\MH^2\times\MV$, where $\MH^2$ is a one-sheet
hyperboloid embedded in three-dimensional Minkowski space (or its
universal covering, resp.) and $\MV$ is a Euclidean surface to be
specified below. This class of vacuum solutions to the Einstein equations
yields solutions with curvature or conical singularities located along
space-like strings or domain walls of curvature singularity evolving in
time. All singularities are naked ones.

Case $C$ is very similar to the spatially symmetric solutions (case $B$)
but has important new features. First, we cannot restrict ourselves to
positive definite metrics $h_{\mu\nu}$ on $\MV$ because one of the
Einstein equations (\ref{escurv}) is not invariant under the transformation
$h_{\mu\nu}\rightarrow-h_{\mu\nu}$ for a fixed $m$. Note that in the
case $k=\const$ the transformation $g_{\al\bt}\rightarrow-g_{\al\bt}$
may be always compensated by exchanging space and time coordinates on
$\MU$, $\tau\leftrightarrow\s$ leaving equation (\ref{escuru}) invariant.
This is impossible for the Euclidean signature metric on $\MV$.
Therefore we fix $m=1$ to have the same signs in the equation as in the
case $k=1$, but allow the metric $h_{\mu\nu}$ to be both positive or
negative definite.

To obtain solutions of equations (\ref{escuru})--(\ref{einnmm}) one may
follow the same steps as before replacing everywhere $m$ by $k$ and
$g_{\al\bt}$ by $h_{\mu\nu}$. Therefore we merely sketch the whole
procedure stressing only particularly important points resulting from
the Euclidean signature. Now the conformal gauge is
\begin{equation}                                        \label{ecogae}
  h_{\mu\nu}dy^\mu dy^\nu=2hdzd\bar z=2h(d\rho^2+d\s^2)
\end{equation}
where $h(z,\bar z)$ is a function of the complex coordinates on $\MV$,
\begin{equation}                                        \label{ecocov}
  z=\rho+i\s,~~~~~~\bar z=\rho-i\s,
\end{equation}
where $\rho=y^2$, and $\s=y^3$. In the total line element
\begin{equation}                                        \label{eincoe}
  ds^2=kd\Phi^2+2hdzd\bar z,
\end{equation}
$d\Phi^2$ is a constant curvature metric on $\MU$ given, for
example, by (\ref{ecoclm}). For the two unknown functions $k$ and $h$
we have, instead of equations (\ref{egcoga})--(\ref{egcogc}),
\begin{eqnarray}                                       \label{egcoha}
  \pl_z\pl_z k -\frac{\pl_zk\pl_zk}{2k}-\frac{\pl_z h\pl_zk}h&=&0,
\\                                                     \label{egcohb}
  \pl_{\bar z}\pl_{\bar z}k -\frac{\pl_{\bar z}k\pl_{\bar z}k}{2k}
  -\frac{\pl_{\bar z}h\pl_{\bar z}k}h&=&0,
\\                                                     \label{egcohc}
  \frac{2\pl_z\pl_{\bar z}k}g-2(k\Lm+K)&=&0.
\end{eqnarray}
The solution of (\ref{egcoha}) and (\ref{egcohb}) is
$k=k(z\pm\bar z)$, $h=h(z\pm\bar z)$, and
\begin{equation}                                        \label{ehintk}
  h=\pm\frac{|k'|}{4\sqrt{|k|}},
\end{equation}
where the upper and lower signs correspond to positive and negative
definite metrics on $\MV$, respectively. Thus, $k$ and $h$ depend on
either $\rho$ or $i\s$. However, due to the rotational symmetry of
(\ref{ecogae}) these two choices are equivalent and we assume for
definitness that $k$ and $h$ are functions of $\rho$.
Then the Laplacian is $4\pl_z\pl_{\bar z}k=k''$ (no $\pm$ sign!). Now
instead of (\ref{esecom}) we get
\begin{equation}                                       \label{esecok}
  k''\mp(k\Lm+K)\frac{|k'|}{\sqrt{|k|}}=0,
\end{equation}
with the same sign conventions as in (\ref{ehintk}). Using the
result that the curvature has a true singularity at $k=0$ we
restrict the range of $k$ to positive values. Then equation (\ref{esecok})
may be integrated to give
\begin{equation}                                       \label{efiokw}
  \left|\frac{dk}{d\rho}\right|=\mp 2W(k),
\end{equation}
where
\begin{eqnarray}                                       \label{efiokp}
  W(k)&=&-\frac13\Lm k^{3/2}-Kk^{1/2}-2M,~~~~~~k>0\,,
\end{eqnarray}
$M$ being an arbitrary integration constant. Although it cannot be
interpreted as a mass we use the same notation as before to simplify
comparison. With the parametrization $k=r^2$ for positive $k$ a general
hyperbolic solution to the Einstein equations is, finally,
\begin{equation}                                        \label{egshty}
  ds^2=r^2d\Phi^2-N(r)(d\rho^2+d\s^2),
\end{equation}
where
\begin{equation}                                        \label{efunhs}
  N(r)=-K-\frac{2M}r-\frac13\Lm r^2,
\end{equation}
and the function $r=r(\rho)$ is defined by the equation
\begin{equation}                                        \label{eforrh}
  \left|\frac{dr}{d\rho}\right|=\mp N(r).
\end{equation}
The modulus sign in this equation means that in each case there are
two solutions $r(\rho)$ and $r(-\rho)$. From equation (\ref{egshty})
it is clear that for $N>0$ and $N<0$ the metric on $\MV$ is negative
and positive definite, respectively. Taking the function $r$ as a
coordinate (\ref{egshty}) can be written in Schwarzschild-like form
\begin{equation}                                        \label{egssty}
  ds^2=r^2d\Phi^2-\frac{dr^2}{N(r)}-N(r)d\s^2.
\end{equation}

The resulting hyperbolic solution has three Killing vector fields
generating the symmetry group $SO(1,2)$ of the one-sheet hyperboloid of
constant curvature and one Killing vector field $\pl_\s$ for the $\MV$
surface.

Calculations similar to the $k=1$ case yield explicit expression for
the scalar curvature on $\MV$,
$$
  R^h=\frac23\Lm+\frac{4M}{r^3},
$$
and for the invariant eigenvalue of the Weyl tensor we get the same
expression as given in case $B$ by (\ref{ectsqu}). This justifies the
range of $k=r^2\in(0,\infty)$.
\subsection{Global structure                           \label{solcot}}
The global structure of a hyperbolic solution (\ref{egshty}) or
(\ref{egssty}) depends on the zeros of $N$:
$$
  N(r_k)=0.
$$
The construction of maximally
extended solutions requires a careful analysis of the extremals for a
given metric. The local solutions of $2d$ gravity with torsion
have the same structure and therefore the analysis performed in
\cite{Katana97} may be applied directly in the present case. To save
space we only formulate the rules of how to construct
maximally extended Euclidean solutions from the Lorentzian ones:
\begin{itemize}
\item Each of the Lorentz signature solution decomposes into disconnected
solutions for the Euclidean metric along horizons. That is each
conformal block by itself represents one maximally extended solution
in the Euclidean case.
\item Homogeneous, $N<0$, and static, $N>0$, conformal blocks represent
surfaces of positive and negative definite metrics, respectively.
\item Each of the horizons shrinks to a point in the Euclidean case and
lies at a finite distance for a simple zero of $N$ and at infinite
distance for a higher zero of $N$.
\item The completeness or incompleteness of curvature singularities is
the same as before.
\end{itemize}

To obtain a maximally extended Euclidean surface $\MV$ one must identify
the points $\s$ and $\s+L$ along the Killing direction \cite{Katana97}
for a simple zero.
The obtained surface is not a cylinder because its circumference shrinks
to zero if one approaches a horizon, that is the zero of $N$. In general,
one gets a conical singularity there with the deficit angle
$$
  \triangle\omega=\frac12 L\left|N'(r_k)\right|-2\pi
$$
where $N'$ denotes the derivative. Hence the necessary
and sufficient condition for the absence of a conical singularity is
\begin{equation}                                        \label{edangr}
  L=\frac{2\pi}{\left|\frac M{r_k^2}-\frac13\Lm r_k\right|}.
\end{equation}
If this equation holds then we get a smooth surface at this point and
the whole space-time may be totally smooth (if there are no curvature
or further conical singularities). For a double zero the points
corresponding to this horizon lie at an infinite distance and one
does not need to bother about conical singularities.
Again the classification of global hyperbolic solutions depends
on the number and type of zeros of $N$ defined by the constants
$K$, $\Lm$, and $M$.
\subsection{Hyperbolic solutions for $K=-1$            \label{solhyp}}
Let us note that the cases $K=1$ and $K=-1$ may be obtained from each
other by permutation of the first two coordinates.
We choose $K=-1$ to retain the same expression for $N$ as for spherically
symmetric solutions. The line element for the one sheet hyperboloid
(\ref{ecoclm}) in angular coordinates reads
$$
  d\Phi^2=d\theta^2-\cosh^2\theta d\f^2,~~~~~~
  -\infty<\theta<\infty\,,~~0\le\f<2\pi \,,
$$
so that the corresponding $4d$ line element becomes
\begin{equation}                                        \label{egshts}
  ds^2=r^2(d\theta^2-\cosh^2\theta d\f^2)-\frac{dr^2}{N(r)}-N(r)d\s^2,
\end{equation}
where $N(r)$ is given by (\ref{edefus}) again.
It has precisely the same form as for the Kottler solution
\cite{Kottle18} but here it enters the Euclidean part of the metric.
Note that now the surface $\MU=\MH^2$ or its universal
covering space is completely smooth. The surface $\MV$ may have a negative
($N>0$) or positive ($N<0$) definite metric. For negative definite metric
the timelike coordinate is $\theta\in(-\infty,\infty)$, and the
three-dimensional space is the product of a circle, $\f\in[0,2\pi)$, and
the surface $\MV$ to be constructed below. If $\MU$ is the universal
covering space of the hyperboloid $\MH^2$, then the space equals
$\MR\times\MV$. The evolution of space in time lasts forever and if
there are singularities, all of them are naked.

For positive definite metrics on $\MV$ the timelike coordinate is
$\f\in[0,2\pi)$, and the space is a product of a line,
$\theta\in(-\infty,\infty)$, with $\MV$. The corresponding space-time
contains closed timelike curves (including extremals) unless
$\MU$ is chosen to be the universal covering of $\MH^2$.
\subsubsection{Minkowskian space-time, $\Lm=0$, $M=0$}
In this case the surface $\MV$ is a half plane $r\in(0,\infty)$,
$\s\in(-\infty,\infty)$. It is incomplete at $r=0$ due to a coordinate
singularity which was also the case for the spherical coordinates.
\subsubsection{ $\Lm=0$, $M>0$}
%
\begin{figure}[t]
 \begin{center}
 \leavevmode
 \epsfxsize 14cm \epsfbox{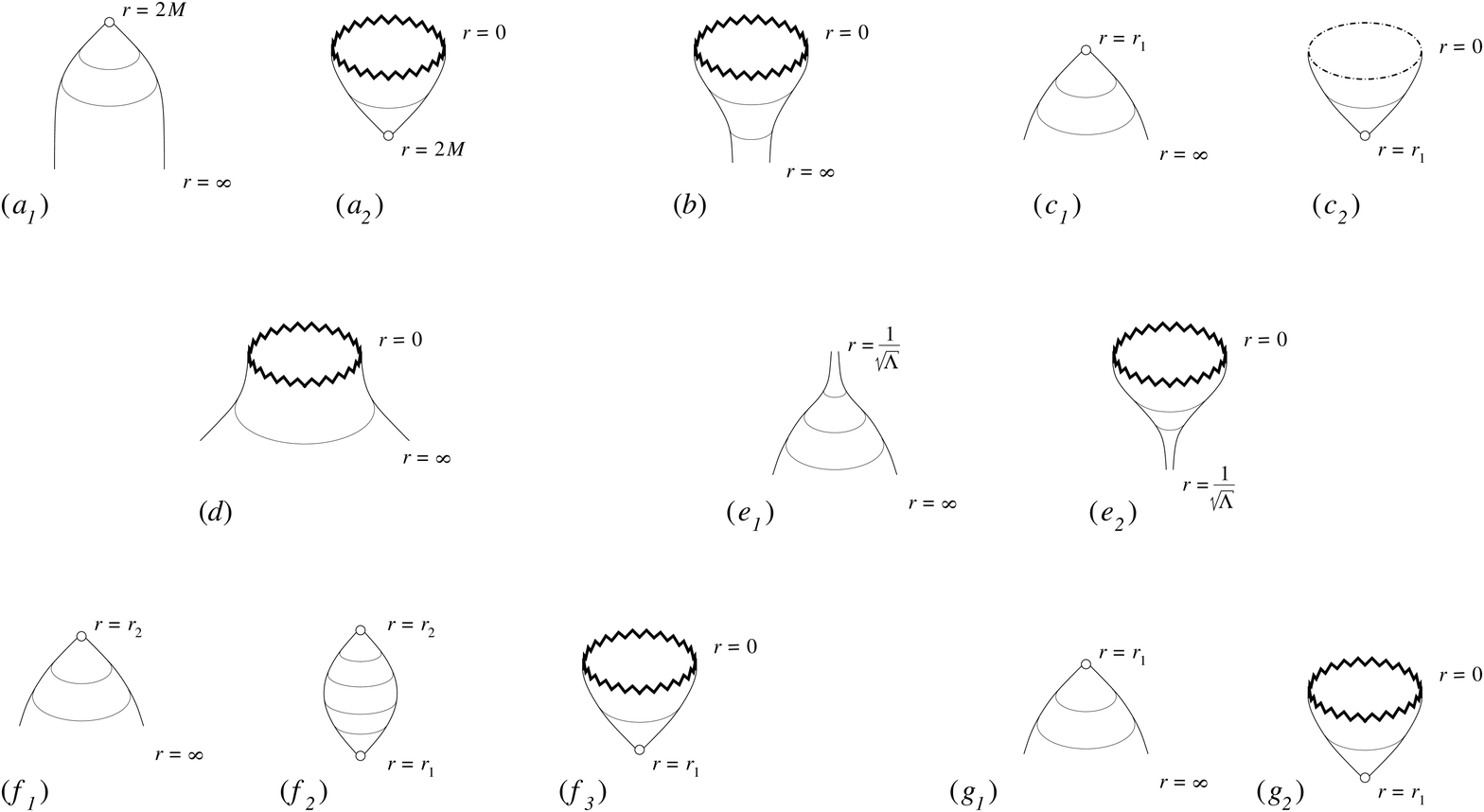}
 \end{center}
 \renewcommand{\baselinestretch}{.9}
 \small \normalsize
 \begin{quote}
 {\bf Figure \FigC:} {\small Hyperbolic solutions for $K=-1$.
         The hollow circles denote possible conical singularities.}
 \end{quote}
\end{figure}
This is the counterpart of the Schwarzschild solution.
There are two different and disconnected Euclidean surfaces shown in
Fig.~{\FigC}~$(a_1)$ and $(a_2)$ with negative and positive definite metric.
There we have identified the points $\s$ and $\s+L$. The
global solution corresponding to the surface $(a_1)$ has the signature
$(+---)$ and is globally smooth or may contain a conical singularity
at $r_*=2M$. Topologically it is a plane and in the case of a conical
singularity the whole space-time describes the infinitely long propagation
of an infinite cosmic string. This solution has no further curvature
singularity and if equation (\ref{edangr}) is fullfiled the whole
space-time is totally smooth. A particular feature of this solution
which is the counterpart of the asymptotic flatness of the Schwarzschild
solution is that the cylinder shown in $(a_1)$ has constant circumference
as $r\rightarrow\infty$. Surface $(a_2)$ corresponds to another global
solution with the signature $(+-++)$. The surface $\MV$ is topologically
a disc with singular curvature at the boundary $r=0$ and, possibly, a
conical singularity at the center. Hence the three-dimensional space is
a cosmic string surrounded by a cylindrical domain wall of singular
curvature. The singular boundary cannot be drawn adequately because
it lies at a finite distance but has infinite circumference.
\subsubsection{ $\Lm=0$, $M<0$}
Here the surface $\MV$ is a half plane $r\in(0,\infty)$,
$\s\in(-\infty,\infty)$ with negative definite metric.
The singularity lies along the line $r=0$.
{}From the four-dimensional point of view it represents an
infinite planar domain wall of singular curvature located at a finite
distance which lives forever.

One may also identify the points $\s$ and $\s+L$. The resulting
surface is shown in Fig.~{\FigC}~$(b)$. The corresponding three-dimensional
space represents a cylindrical domain wall.
\subsubsection{De Sitter solution $\Lm>0$, $M=0$}
This is one more version of the de Sitter solution. There are two surfaces
shown in Fig.~{\FigC}~$(c_{1,2})$ with positive and negative definite metrics,
respectively. They are obtained after identifying $\s$ with
$\s+L$. The de Sitter solution corresponds to their universal covering
spaces. The surface $(c_2)$ has no singularity at $r=0$ but is
incomplete due to the coordinate choice. The present form shows how the
de~Sitter solution can be deformed to obtain a conical singularity at
$r_1=\sqrt{3/\Lm}$.
\subsubsection{Anti de Sitter solution $\Lm<0$, $M=0$}
In the anti de Sitter case the $\MV$ surface is a half plane
$r\in(0,\infty)$, $\s\in(-\infty,\infty)$ with negative definite
metric. It has no singularity but is incomplete at $r=0$.
\subsubsection{              $\Lm>0$, $3M>\frac1{\protect\sqrt{\Lm}}$}
The surface $\MV$ is the half plane $r>0$ with positive definite
metric and a curvature singularity at $r=0$. It describes a planar
domain wall. Compactification of $\s$ yields the exterior
region of a circular domain wall shown in Fig.~{\FigC}~$(d)$.
\subsubsection{              $\Lm>0$, $3M=\frac1{\protect\sqrt{\Lm}}$}
For a double root horizon two surfaces shown in Fig.~{\FigC} $(e_{1,2})$
are obtained, both with negative definite metric. The surface $(e_1)$ is
not simply connected. Topologically it is a plane with its centre removed
to infinity.
The surface $(e_2)$ corresponds to a cylindrical domain wall
of singular curvature.
The axis of the cylinder is located at an infinite distance.
\subsubsection{            $\Lm>0$, $0<3M<\frac1{\protect\sqrt{\Lm}}$}
Due to the two zeros $r_{1,2}$ we obtain three maximally extended
solutions, Fig.~{\FigC} $(f_{1-3})$.
The surfaces $(f_1)$ and $(f_3)$ were qualitatively described above
but the surface $(f_2)$ is new. It is compact and may contain two conical
singularities. An algebraic analysis of the system of equations
$N(r_1)=N(r_2)=0$ and $N'(r_1)=-N'(r_2)$ with $r_1\ne r_2$ shows that
it has no positive roots. Therefore, by fitting appropriately the period $L$
of compactification we can avoid one of them but not both simultaneously,
so there exists necessarily a cosmic string. Topologically, the space is a
product of a sphere with a line where the sphere must contain at least one
conical singularity.
\subsubsection{$\Lm>0$, $M<0$}
The function $N$ has one zero so we obtain two Euclidean surfaces. The first
$(g_1)$ has the same form as $(c_1)$, and the solution describes an
infinite cosmic string without curvature singularity. The second surface
$(g_2)$ has the form shown in $(f_3)$ but with negative definite metric.
\subsubsection{ $\Lm<0$, $M>0$}
This case has two inextendible surfaces for $\MV$ as in the previous
case $(g_{1,2})$, but the signature of the metric on both surfaces must
be changed.
\subsubsection{ $\Lm<0$, $M<0$}
This case is similar to the case $(d)$ $\Lm>0$, $3M>1/\sqrt\Lm$,
but with negative definite metric.
\subsection{Minkowski plane solutions, $K=0$           \label{spmprp}}
For $K=0$ the surface $\MU$ is the Minkowskian plane, a cylinder, or
a torus and some new topologically interesting solutions arise.
The corresponding line element in Schwarzschild coordinates is
\begin{equation}                                        \label{einmps}
  ds^2=r^2(dt^2-dx^2)-\frac{dr^2}{N(r)}-N(r)d\s^2,
\end{equation}
where
$$
  N(r)=-\frac{2M}r-\frac13\Lm r^2.
$$
There are four qualitatively different cases:
\subsubsection{$\Lm=0$, $M\ne0$}
%
\begin{figure}[t]
 \begin{center}
 \leavevmode
 \epsfxsize 7cm \epsfbox{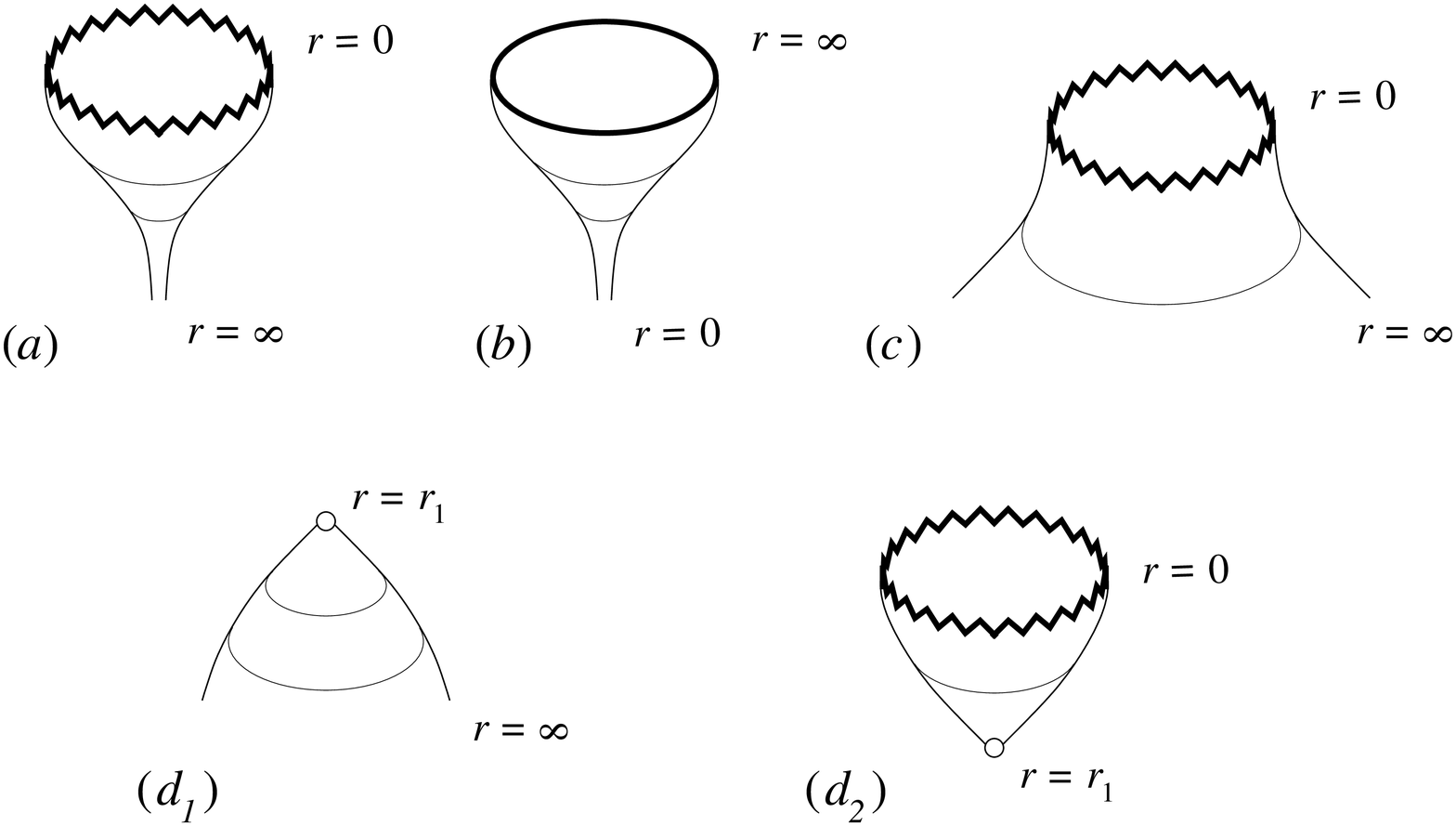}
 \end{center}
 \renewcommand{\baselinestretch}{.9}
 \small \normalsize
 \begin{quote}
 {\bf Figure \FigD:} {\small Minkowski plane solutions, $K=0$.}
 \end{quote}
\end{figure}
In this case we have a surface shown in Fig.~{\FigD}~$(a)$ with positive
and negative definite metric for $M>0$ and $M<0$, respectively.
\subsubsection{(Anti) de Sitter solution $\Lm\ne0$, $M=0$}
This is one more representation of the (anti) de Sitter solution for
positive and negative cosmological constant. The surface $\MV$ is
shown in Fig.~{\FigD}~$(b)$. In this form it is explicitly conformally flat
\begin{equation}                                        \label{edescm}
  ds^2=\frac9{\Lm^2\rho^2}(dt^2-dx^2)+\frac3{\Lm\rho^2}(d\rho^2+d\s^2).
\end{equation}
\subsubsection{$\Lm>0$, $M>0$}
In this case $N$ has no zero, and the surface $\MV$ is shown in
Fig.~{\FigD}~$(c)$.
It describes the outside region of a cylindrical domain wall of
curvature singularity. For $\Lm<0$ and $M<0$ one has simply to change
the signature of the whole metric.
\subsubsection{$\Lm>0$, $M<0$}
With one zero of $N$ the two surfaces shown in Fig.~{\FigD}~$(d_{1,2})$
are obtained. They describe a cosmic string and a cosmic string surrounded
by a domain wall of curvature singularity.
\section{Summary and Outlook                           \label{summar}}
The analysis of global solutions given in the present paper exhausts
all vacuum solutions to the Einstein equations having the form of a warped
product of two surfaces. We have shown that the requirement of maximal
extension almost uniquely determines their causal structure. The solutions
are classified by the values of a cosmological constant, the
constant scalar curvature of one of the surfaces, and an integration
constant which for the Schwarzschild solution coincides with its mass.
Although these solutions have a simple form and were known locally,
their global structure is of great physical interest, describing, e.g.,
cosmic strings, domain walls, cosmic strings surrounded by domain walls,
and solutions with closed timelike curves.
Our analysis becomes possible thanks to a systematic method to construct
maximally extended surfaces which has been developed for $2d$
gravity models. A similar approach may be used also in the more general
problem including matter fields subject to some symmetry restrictions.
The problem seems to be solvable at least in the presence of a $U(1)$ gauge
field. Although the Carter-Penrose diagrams have been presented here for all
cases, we have to defer a systematic discussion of possible special properties
of non-radial geodesics. However, in analogy to the Schwarzschild case we do
not expect any surprises regarding the extendability of $4d$ solutions.
\section*{Acknowledgement}
We thank H.~Balasin and I.~Volovich for discussions. This work has been
supported by the Austrian Fonds zur F\"orderung der wissenschaftlichen
For\-schung (FWF) Project No.\ P 10221-PHY.  One of the authors (M.~K.)
thanks the Austrian Academy of Sciences, the Erwin Schr\"odinger
International Institute for Mathematical Physics,
the Technical University of Vienna,
and the Russian Foundation for Basic Research, Grants RFBR-96-010-0312
and RFBR-96-15-96131, for financial support.

\end{document}